\documentclass[12pt]{emulateapj}

\usepackage{graphicx}
\usepackage{subfigure}
\usepackage{rotating}
\usepackage{lscape}

\newcommand{\eg}{e.g.\@~}

\newcommand{\fig}{Fig.~}
\newcommand{\tbl}{Table~}

\newcommand{\feii}{[Fe \textsc{ii}]}
\newcommand{\sivi}{[Si \textsc{vi}]}
\newcommand{\hei}{He \textsc{i}}
\newcommand{\brg}{Br$\gamma$}
\newcommand{\brd}{Br$\delta$}
\newcommand{\molhy}{H$_2$}

\begin{document}

  \title{The Inner Kiloparsec of Mrk 273 with Keck Adaptive Optics} 
  
  \author{Vivian U\altaffilmark{1,9},
Anne Medling\altaffilmark{2},
David Sanders\altaffilmark{1}, 
Claire Max\altaffilmark{2},
Lee Armus\altaffilmark{3}, 
Kazushi Iwasawa\altaffilmark{4}, 
Aaron Evans\altaffilmark{5,6},
Lisa Kewley\altaffilmark{1,7},
Giovanni Fazio\altaffilmark{8}
}
 
 \altaffiltext{1}{Institute for Astronomy, University of Hawaii,
   2680 Woodlawn Dr., Honolulu, HI 96822, USA}
  \altaffiltext{2}{Department of Astronomy and Astrophysics,
    University of California, Santa Cruz, 1156 High Street, Santa
    Cruz, CA 95064, USA}
  \altaffiltext{3}{Spitzer Science Center, California Institute of
    Technology, 1200 E. California Blvd., Pasadena, CA 91125, USA}
  \altaffiltext{4}{ICREA and Institut del Ci\`{e}ncies del Cosmos,
    Universitat de Barcelona (IEEC-UB), Mart\'{\i} i Franqu\`{e}s, 1,
    E-08028 Barcelona, Spain}
  \altaffiltext{5}{Department of Astronomy, University of Virginia,
    530 McCormick Road, Charlottesville, VA 22904, USA}
  \altaffiltext{6}{National Radio Astronomy Observatory, 520 Edgemont
    Road, Charlottesville, VA 22903, USA}
  \altaffiltext{7}{Research School for Astronomy and Astrophysics,
    Australian National University, Cotter Road, Weston Creek, ACT
    2611, Australia}
 \altaffiltext{8}{Harvard-Smithsonian Center for Astrophysics,
   60 Garden St., Cambridge, MA 02138, USA}
 \altaffiltext{9}{\emph{Current Address:} Department of Physics and
   Astronomy, University of California, Riverside, 900 University
   Avenue, Riverside, CA 92521, USA; vivianu@ucr.edu}
  
  \begin{abstract}
There is X-ray, optical, and mid-infrared imaging and spectroscopic evidence that
the late-stage ultraluminous infrared galaxy merger Mrk 273 hosts a
powerful AGN. However, the exact location of the AGN and the nature of
the nuclei have been difficult to determine due to dust obscuration
and the limited wavelength coverage of available high-resolution
data. Here we present near-infrared integral-field spectra and images
of the nuclear region of Mrk 273 taken with OSIRIS and NIRC2 on the
Keck II Telescope with laser guide star adaptive optics.  We observe
three spatially-resolved components, and analyze the local molecular
and ionized gas emission lines and their kinematics. We confirm the
presence of the hard 
X-ray AGN in the southwest nucleus.  In the north nucleus, we find a
strongly-rotating gas disk whose kinematics indicate a central black
hole of mass  $1.04 \pm 0.1 \times 10^{9} M_{\sun}$.
The \molhy~emission line shows an increase in velocity dispersion
along the minor axis in both directions, and an increased flux with
negative velocities in the southeast direction; this provides direct
evidence for a collimated molecular outflow along the axis of
rotation of the disk.  The third spatially-distinct component appears
to the southeast, 640 and 750 pc from the north and southwest nuclei,
respectively.  This component is faint in continuum emission but shows
several strong emission line features, including [Si VI] 1.964 $\mu$m
which traces an extended coronal line region. 
The geometry of the [Si VI] emission combined with shock models and
energy arguments suggest that [Si VI] in the southeast component must
be at least partly ionized by the SW AGN or a putative AGN in the
northern disk, either through photoionization or through shock-heating
from strong AGN- and circumnuclear starburst-driven outflows. This
lends support to a scenario in which Mrk 273 may be a dual AGN system.  

  \end{abstract}
  

  \section{Introduction}
  \label{Introduction}
When gas-rich galaxies collide and merge, gaseous material can lose
angular momentum and funnel toward the center of the merger
system~\cite[\eg][]{Barnes92}. There it can feed starburst and active
galactic nuclei (AGN) activities which power luminous and ultraluminous
infrared galaxies ((U)LIRGs).  The detailed processes of how merger
dynamics contribute to AGN
activity~\cite[][]{Sanders88,Hopkins06,Comerford09}, how mass  
builds up  in the central supermassive black
hole~\cite[][]{Hopkins10,Treister10,Rosario11}, how much energy is
deposited into the system by the ionizing
sources~\cite[][]{Younger09,Sanders96}, and when the active nuclei are
triggered~\cite[][]{Donley10,Hopkins12,Farrah09} are areas of active
research.  Because AGN have been observed in mergers with double
nuclei~\cite[][]{Komossa03,Ballo04,Koss12,Fu11,Liu11,McGurk11,Gerke07,Mazzarella12}, 
it is known that an AGN can ``turn on'' before the final coalescence
of the nuclei.  However, many mergers do not host dual
AGN~\cite[e.g.][]{Teng12}.  It is inconclusive from the models when an
AGN is triggered and what dictates whether or not synchronous ignition
occurs~\cite[][]{vanWassenhove12}. 

These questions can be addressed with high-resolution
observations of the nuclei in nearby luminous infrared galaxy mergers.  
Previous integral-field spectroscopic work on large samples of
(U)LIRGs have mapped the large-scale distribution and kinematics of
ionized gas, compared them to that of stellar components,
determined metallicity gradients, and found evidence for galactic
outflows and shocks~\cite[\eg][]{Westmoquette12,Rich11,Rich12,Garcia-Marin09a}.  
Much of the gas is concentrated within the dusty cores of
these (U)LIRGs, so longer wavelengths must be probed to understand the
nuclear regions. ~\cite{PiquerasLopez12} has presented an atlas of the
2D molecular and ionized gas structure in the southern local (U)LIRGs
as probed in the near-infrared, showing the extent and kinematics of the
gaseous components at $\sim$0.2 and 0.9 kpc spatial resolution. These
seeing-limited observations with an average angular resolution 
of $\sim 0\farcs63$ full-width half-maximum (FWHM) set the framework
of the large kiloparsec-scale conditions of the gas and prompt detailed
high-resolution investigations of the nuclear regions.

Due to the advent of laser guide star adaptive optics~\cite[LGS
AO;][]{Wizinowich06,vanDam06}, black holes in local (U)LIRGs can
now be examined at the subarcsecond
level~\cite[][]{Max07,Melbourne08,Engel10,Davies10,Medling11}. 
In particular, the OH-Suppressing Infra-Red Imaging
Spectrograph~\cite[OSIRIS;][]{Larkin06}, an integral-field unit (IFU) with
near-infrared wavelength coverage (1$\mu$m $< \lambda <$ 2.4$\mu$m) on the 
Keck II telescope, offers imaging and kinematic information at $< 
0\farcs1$ spatial resolution with LGS
AO~\cite[\eg][]{Wright09,Law09,Do09,McConnell11,Walsh12}.  NIRC2 (PIs:
Keith Matthews and Tom Soifer), a diffraction-limited infrared camera
also on the Keck II telescope, offers scales of $\sim0.01"$/pixel in the
$1-5 \mu m$ range in its narrow camera mode.  
They both probe the near-infrared regime where dust extinction does
not pose as severe a problem. 

\begin{figure*}
  \centering
  \includegraphics[width=0.9\textwidth]{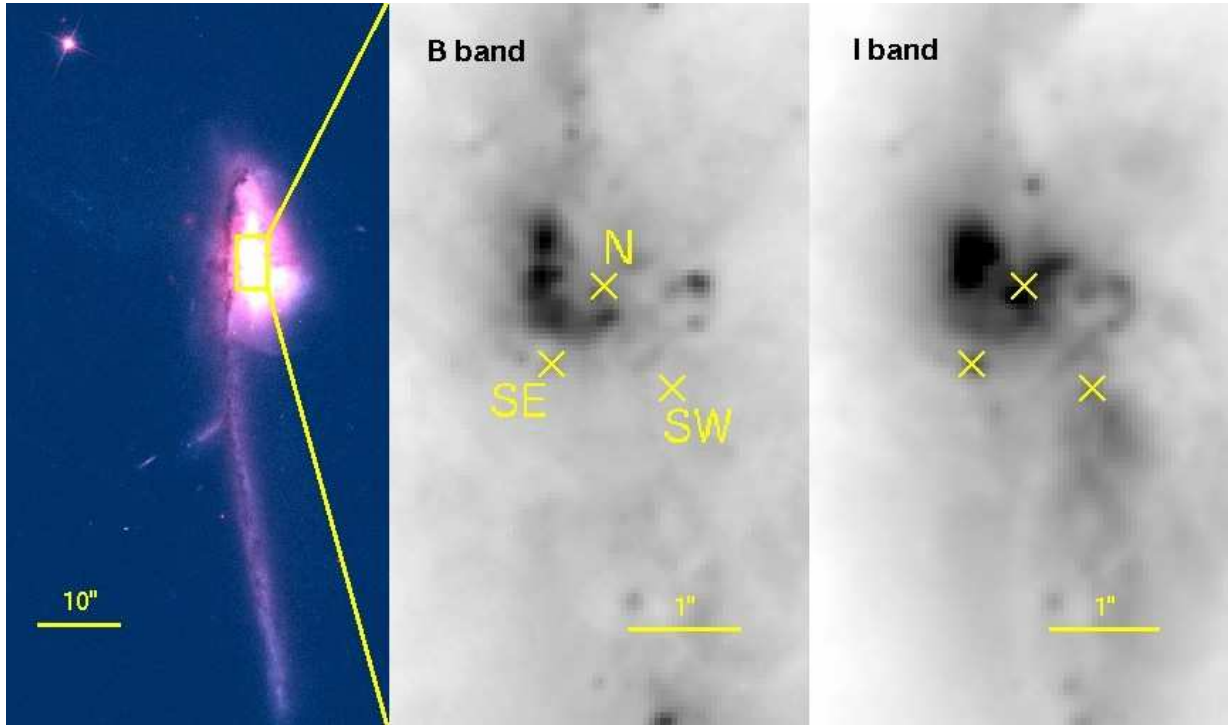}
  \caption{(Left) \emph{HST} ACS $B (F435W) + I (F814W)$ composite
    image of Mrk 273. In these and subsequent images, north is up and
    east is to the left. The box indicates the zoomed in region in the
    grey-scale panels. (Middle) and (right) are the zoomed in $B$- and
    $I$-band images, respectively. The three crosses mark the N, SE,
    and SW sources of interest identified in radio observations
    \cite[][]{Carilli00,Bondi05} whose near-infrared characteristics will be
    presented in this paper.} 
  \label{fig:hst_overview}
\end{figure*}

To understand the mechanisms responsible for powering the extreme infrared
luminosities in (U)LIRGs, we are conducting a survey of nearby galaxy
mergers using OSIRIS to probe the gas kinematics and energetics of the
nuclear regions in these systems. Observed as part of the larger
campaign, Mrk 273 (= UGC 08696) is the focus of this paper as an
advanced merger with infrared luminosity $L_{\rm IR} = 10^{12.21}
L_\odot$. At a distance of $z$ = 0.038 (systemic velocity $cz =
11,400$ km s$^{-1}$; physical scale = 0.754
kpc/\arcsec)\footnote{Physical scale has been computed with a cosmology
calculator~\cite[][]{Wright06}.}, it
exhibits a prominent optical tidal tail ($\sim$40 kpc) extending to
the south (\fig \ref{fig:hst_overview}). With its high infrared
luminosity, this ULIRG has a heated dusty core that obscures the
central ionizing sources in optical light.  As one of the brightest
and closest ULIRGs, Mrk 273 has been well-studied across the full
wavelength spectrum, including \emph{Spitzer}, \emph{Hubble},
\emph{Chandra}, and \emph{GALEX} observations from the Great
Observatories All-sky LIRG
Survey~\cite[GOALS;][]{Armus09,Howell10,Iwasawa11_cgoals}. However,
the details of its power engine remain controversial. 

There is substantial evidence in the literature suggesting that at
least one AGN resides within Mrk 273:~\cite{Armus07} detected the [Ne
V] line in \emph{Spitzer's} mid-infrared IRS spectrum, and measured
hot gas with $T > 300$K in the continuum. Polycyclic aromatic
hydrocarbon strength diagnostics from mid-infrared Infrared Space
Observatory spectroscopy classify the galaxy as an 
AGN~\cite[][]{Lutz99,Rigopoulou99}.  Hard X-ray emission and the Fe K
line~\cite[e.g.][]{Xia02,Iwasawa11_mrk273,Ptak03} have also been
detected in support of the presence of a hard X-ray AGN, though most
of the total power is likely to be from circumnuclear
starbursts~\cite[][]{Iwasawa11_mrk273}.  

The status of Mrk 273 as a ULIRG and a late-stage pre-coalescence
merger makes it an important galaxy for understanding the role of
mergers in galaxy evolution.  The specific goals of our
high-resolution IFU study are to locate the nuclei of the progenitor
galaxies, to determine the characteristics of any supermassive black holes, to verify
their active or dormant nature, and to investigate the relative effect
of AGN activity and star formation on the surrounding nuclear regions.

Multiwavelength studies have decomposed the nuclear region into
multiple components. Three main
components have been identified from the detected radio continuum:
north, southeast, and southwest~\cite[N, SE,
SW;][]{Condon91,Smith98} as indicated in \fig
\ref{fig:hst_overview}. These components show dissimilar
characteristics in most bands; it is not clear which of them may be 
remnant nuclei.  We combine our data with existing evidence to
determine the likely sources of each.  The SE emission is only weakly
detected in the near-infrared \cite[][]{Scoville00} and weak or
undetectable in the X-ray given the limited resolution of
\emph{Chandra}~\cite[][]{Iwasawa11_mrk273}, but the N and SW X-ray emission 
coincide with peaks in the near-infrared continuum flux.  OH maser emission
along with steep radio spectral index suggest the presence of a
low-luminosity AGN with a binding mass of 1.39$\pm$0.16 $\times 10^9
M_\odot$ in the N component~\cite[][]{Klockner04}.  The N radio-1.4GHz
emission is further  resolved into individual compact sources thought
to be clustered supernovae~\cite[][]{Smith98,Carilli00,Bondi05}. 
More recently, \cite{Iwasawa11_mrk273} found, in \emph{Chandra} images
with improved astrometry, that the hard X-ray flux is superposed on
the SW compact source, with a 6-7keV excess extending 
toward the north. They determined that the known hard X-ray AGN is
located at the SW component, and attributed the 6$-$7 keV excess to
enhanced Fe K emission either from the SW source or from a heavily
absorbed Compton-thick AGN at the N component.

In this paper, we present high-resolution infrared integral-field
spectroscopy and imaging observations of Mrk 273 taken with the OSIRIS
and NIRC2 instruments on the Keck II Telescope, probing down to the $\leq
0\farcs1$ scale. These near-infrared high-resolution maps provide
great detail for identifying the location of the hardest photons from
AGN and shocks as well as the dynamics of the gas. Observations and
data processing are described in \S 
\ref{Observations}. Resulting spectra and images are analyzed and
presented in \S \ref{Analysis} and \S \ref{Results}; our
interpretations are detailed in \S \ref{Discussion}.  Our conclusions
are summarized in \S \ref{Summary}. We adopt $H_0 =
70$\,km\,s$^{-1}$\,Mpc$^{-1}$, $\Omega_{\rm m}$ = 0.28, and
$\Omega_\Lambda$ = 0.72 throughout the paper.

  \section{Observations and Data Reduction}
  \label{Observations}

\subsection{OSIRIS Data Cubes}
The Keck OSIRIS observations were obtained on UT 2011 May 22 under
clear, dry, and stable conditions with the LGS AO system.  Two
filter-scale combinations were used to observe Mrk 273: the broad
$K$-band filter ($\lambda = 1965-2381$nm; hereafter Kcb) at the
100mas/pixel scale ($\sim$75pc for Mrk 273), and a narrow $H$-band
filter ($\lambda = 1652-1737$nm; hereafter Hn4) at the 35mas/pixel
scale ($\sim$26pc for Mrk 273).  The tip-tilt
star employed for the AO corrections 
had $R$ magnitude $m_R = 16.1$ at $33\arcsec$ away from the
target, well within the LGS requirement constraints.  We adopted a
position angle of 30$^\circ$ East of North and an observing scheme of
object-sky-object with 10-minute integration per frame, totaling
50-minute and 60-minute on-source times for the Kcb-100mas and
Hn4-35mas observations, respectively.  The A0V star 81 UMa was
observed at 2 different airmasses in the Kcb filter (3 in Hn4) for telluric corrections. 

The data sets were processed using the OSIRIS pipeline version
2.3~\cite[][]{Krabbe04}. The full Astronomical Reduction Pipeline template 
(ARP\_SPEC) incorporates
dark-frame subtraction, channel level adjustment, crosstalk removal,
glitch identification, cosmic ray cleaning, spectra extraction for
data cube assembly, dispersion correction, scaled sky subtraction for
enhanced OH-line suppression, and telluric correction. The extracted
spectra were subsequently cleaned to remove bad pixels. The cleaned
frames were registered using the SW peak position and combined  
using the pipeline's clipped mean algorithm.
The final data cubes encompass signal at each pixel in the $x$, $y$,
and wavelength dimensions. The FWHM of the
tip-tilt star was 0$\farcs$05 in the Kcb setting and 0$\farcs$06 in
the Hn4 setting. 

The final OSIRIS Kcb and Hn4 cubes have been flux-calibrated to the
respective $K'$- and $H$-band NIRC2 flux-calibrated images (see \S
\ref{sec:nirc2}) in the following manner.   Spatially-coincident
regions were first defined within resolution uncertainties in the
OSIRIS and NIRC2 images of the matching bands.  Spectroscopically, the
bandwidth differences between the OSIRIS data cubes and the NIRC2
images have been taken into account while integrating over the Kcb and
Hn4 wavelength ranges, respectively.  Conversion factors thus
determined offer a calibration to absolute fluxes in physical units
that allow for line ratio computations across the two bands.  

It is important to understand the point-spread function (PSF) intrinsic
to our OSIRIS observations before comparing them to models.  We estimate 
our PSF by fitting a Moffat profile to images of the tip-tilt star
taken right before our galaxy observation sets.  However, this is a
best-case limit for the PSF, because the AO correction will degrade
somewhat when 
moving off-axis to the galaxy.  The tip-tilt star provides low-order
corrections, which vary on scales of the isokinetic angle.  (This
angle is much larger than the isoplanatic angle, which governs how
wide a field-of-view a laser guide star can correct.)  At Mauna Kea,
the isokinetic angle is approximately 75" \cite[][]{vanDam06}; this 
represents the tip-tilt star distance at which Strehl will be degraded
by $1/e$.  Because our tip-tilt star is at a separation of $\sim33$"
from Mrk 273, we expect our Strehl to be degraded by a factor of
$\sim0.6$ from the tip-tilt star profile.  To model this, we convolve
our tip-tilt star PSF model by a Gaussian of FWHM 0$\farcs$05 to
estimate our true PSF. 

\subsection{NIRC2 Images}
\label{sec:nirc2}
High-resolution images were also obtained with NIRC2, using the
narrow camera mode (0$\farcs$009942 pixel$^{-1}$; 8 pc pixel$^{-1}$)
and the LGS AO System, on UT 2012 May 20.  Images were taken in the $H$-
($\lambda_{c} = 1633$ nm) and $K'$-($\lambda_{c}$ = 2124 nm) bands, with
exposure times of 9 minutes each. Individual frames were exposed for
30 seconds for each of two coadds; frames were dithered using the
standard `bxy3' dither pattern that avoids the noisy lower-left
quadrant of the chip. The same tip-tilt star used in the OSIRIS
observations was used here.  We reduced our images using the CfAO
Treasury Survey pipeline~\cite[][]{Glassman02,Melbourne05}.  A sky
frame was produced by combining sky regions of each dithered frame,
then subtracted from each frame.  A flat field was created by
normalizing the constructed sky frame. This flat field was then
divided out of each frame in order to correct for varying
sensitivities across the detector.  The sky-subtracted-and-flattened
images were subsequently aligned and median-combined to obtain the final
reduced image, which were then flux-calibrated using images taken 
of the UKIRT Faint Standard star FS~133 \cite[][]{Hawarden01}.  Standard
star exposures were 5 seconds per frame, also dithered with the `bxy3' script; 
four frames were combined for the $H$-band standard, three 
for the $K'$-band standard. A FWHM of $\sim$0$\farcs$06 has been 
measured from a compact source south of the SE component in both 
the $H$- and $K'$-band images.

  \section{Emission Line Fitting and Error Analysis}
  \label{Analysis}
  
From each of the reduced OSIRIS Kcb and Hn4 data cubes, various
emission lines were mapped as follows.  First we extracted the 
continuum with power-law fits and subtracted it from the spectrum. We
measured the signal-to-noise (S/N) ratio of the lines of interest
based on the line flux and root-mean-square (rms) residuals of the
continuum.  We fit the lines with a Gaussian profile to determine the
flux, velocity (shift from the systemic velocity, 11,400 km s$^{-1}$),
and velocity dispersion (sigma of the Gaussian).   
In order to gauge the accuracy of the line-fit parameters, synthetic
data cubes were then generated using the line parameters fitted to each
pixel. The S/N ratio measured above determined the random noise level
added to the synthetic cube.  We then refit the line parameters of the
synthetic cube 500 times with random noise added.  The final
uncertainty in each parameter's original measurement was obtained from
the distribution of that of the resulting models.

\begin{figure*}
  \centering
  \includegraphics[width=0.9\textwidth]{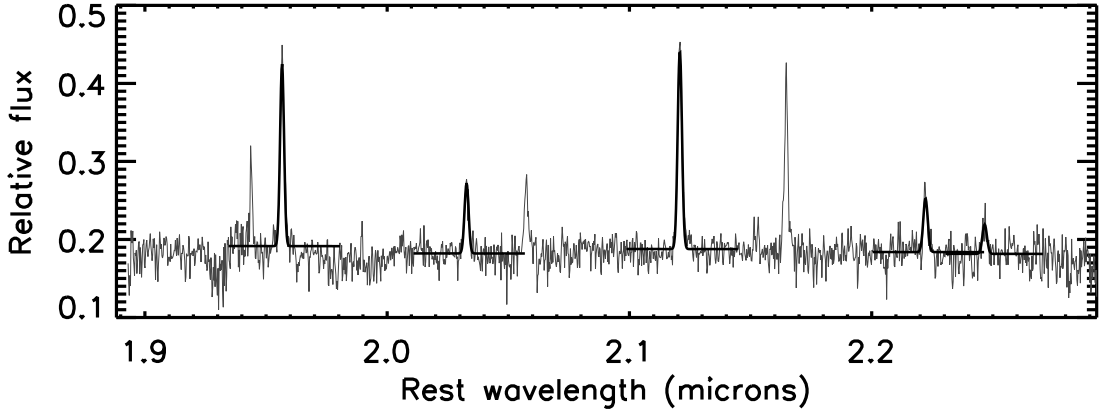}
  \caption{Simultaneous fitting of the different \molhy~transitions
    for an arbitrarily-selected pixel as detected in the Kcb data
    cube.  The thin grey line represents the data spectrum; the thick
    black lines represent the model fits.  For each spectrum, all
    \molhy~lines are required to have the same velocity and velocity
    dispersion; flux is fitted separately for each line.  A similar
    fitting routine is used for \brg~and \brd.}   
  \label{fig:h2linefit}
\end{figure*}

\begin{figure*}
  \centering
  \includegraphics[width=0.9\textwidth]{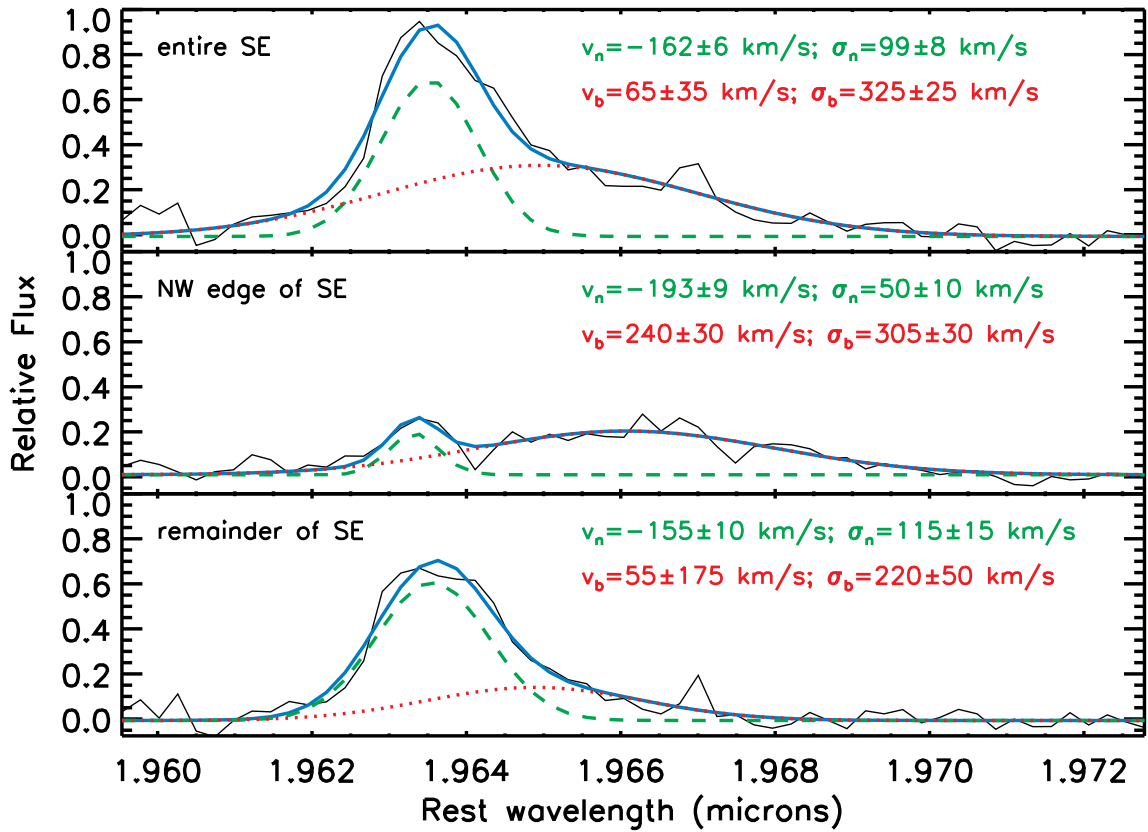}
  \caption{A double-Gaussian component-fit to the \sivi~line profile
    in three regions of the integrated SE component. The top panel includes
    the entire SE component; while the middle and bottom panels
    split the spectrum into the northwest edge and remaining pixels.
    (These regions are identified in the middle panel of
    \fig\ref{fig:binned_sivi}.) 
    In each panel, the black line represents the data spectrum; the
    green dashed and red dotted lines represent the two components
    that contribute toward the total model fit in solid blue.  In the
    total spectrum, 
    the \sivi~line clearly shows two velocity components: one with a narrow 
    velocity dispersion and one with a broader velocity dispersion.  We find
    by splitting the spectrum up spatially that the broader velocity component 
    comes mainly from the northwest edge of the SE component, while the narrow
    portion of the line appears more heavily in the remaining
    section. Parameters for each fitted component are listed in each panel.}   
  \label{fig:twocomp}
\end{figure*}

We adopted this procedure for each line species.  Different lines are
treated separately from each other, but all the \molhy~transitions or
both the \brg~and \brd~lines  are fitted simultaneously.  An 
example of the simultaneous fit of the \molhy~transitions for a typical
pixel in the Kcb cube is shown in \fig \ref{fig:h2linefit}. Visual
inspection ensured that the line fits were sensible with appropriate
continuum levels determined within each spectral window.  
We noted that in many of the pixels with \sivi~detection, the fit of a
single Gaussian profile resulted in large residuals.  In such cases we
fit a broad and a narrow Gaussian component to these lines; this
significantly improved the residuals (\fig \ref{fig:twocomp}). The
pixels associated with the two components appeared to be spatially
segregated. The physical implication of this double-component fit is
further discussed in \S \ref{Other}.

In order to increase the S/N ratio for regions with inadequate flux,
we adopted optimal Voronoi binning for the investigation of the
kinematics using the~\cite{Cappellari03}
code. This algorithm bins pixels together in a region until the emission 
line in that region achieves sufficient S/N, which we chose here to be
3. This allows us to explore the 
spatial resolution in areas with high S/N and to examine those regions
with low S/N in meaningful spatial regions.

  \section{Maps and Data Cubes}
  \label{Results}
  
    \subsection{NIRC2 Near-infrared Images}

\begin{figure*}
  \centering
  \includegraphics[width=0.7\textwidth,angle=90]{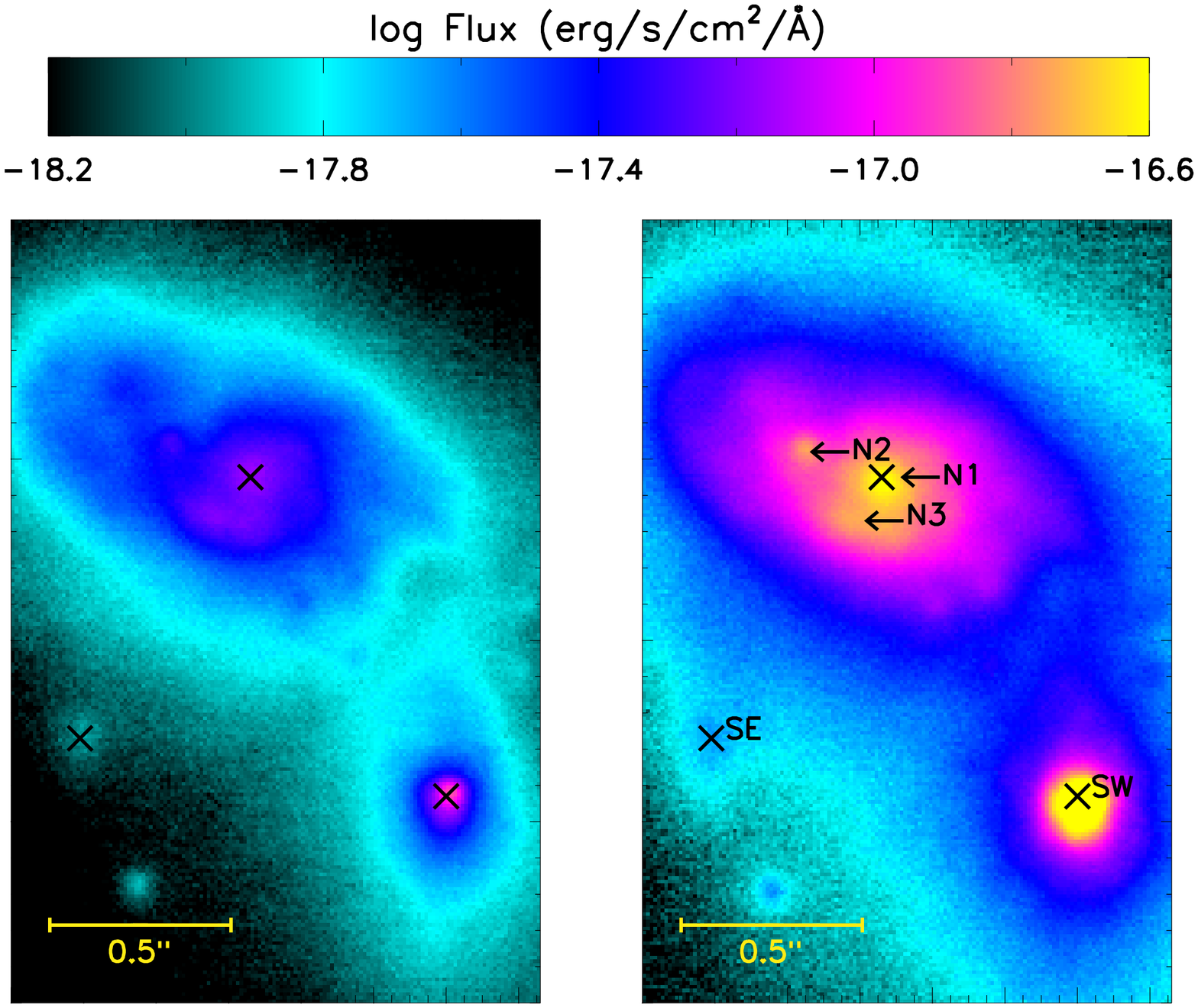}
  \caption{Keck NIRC2 $H$- (left) and $K'$-band (right) images taken
    with the narrow camera (0$\farcs$01 pixel scale). The two images
    are displayed in logarithmic scale in units of
    erg/s/cm$^2$/\AA~with N pointing up. These images reveal clumpy
    structure and wispy gas in the nuclear regions of Mrk 273.
    Crosses denote the three components from
    \fig\ref{fig:hst_overview}, and the arrows point to three
    different clumps resolved in the N component.  See text for a
    detailed discussion of each.}  
  \label{fig:nirc2}
\end{figure*}

Taken at the $0\farcs01$-per-pixel narrow-camera mode, the $H$- and
$K'$-band NIRC2 images provide an unprecedented high-resolution view
of the nuclear components in these wavelengths (\fig
\ref{fig:nirc2}). All three components, as denoted by the crosses
corresponding to the N, SW, and SE, are distinct in both bands.  

\begin{figure*}
  \centering
  \includegraphics[width=0.85\textwidth,angle=90]{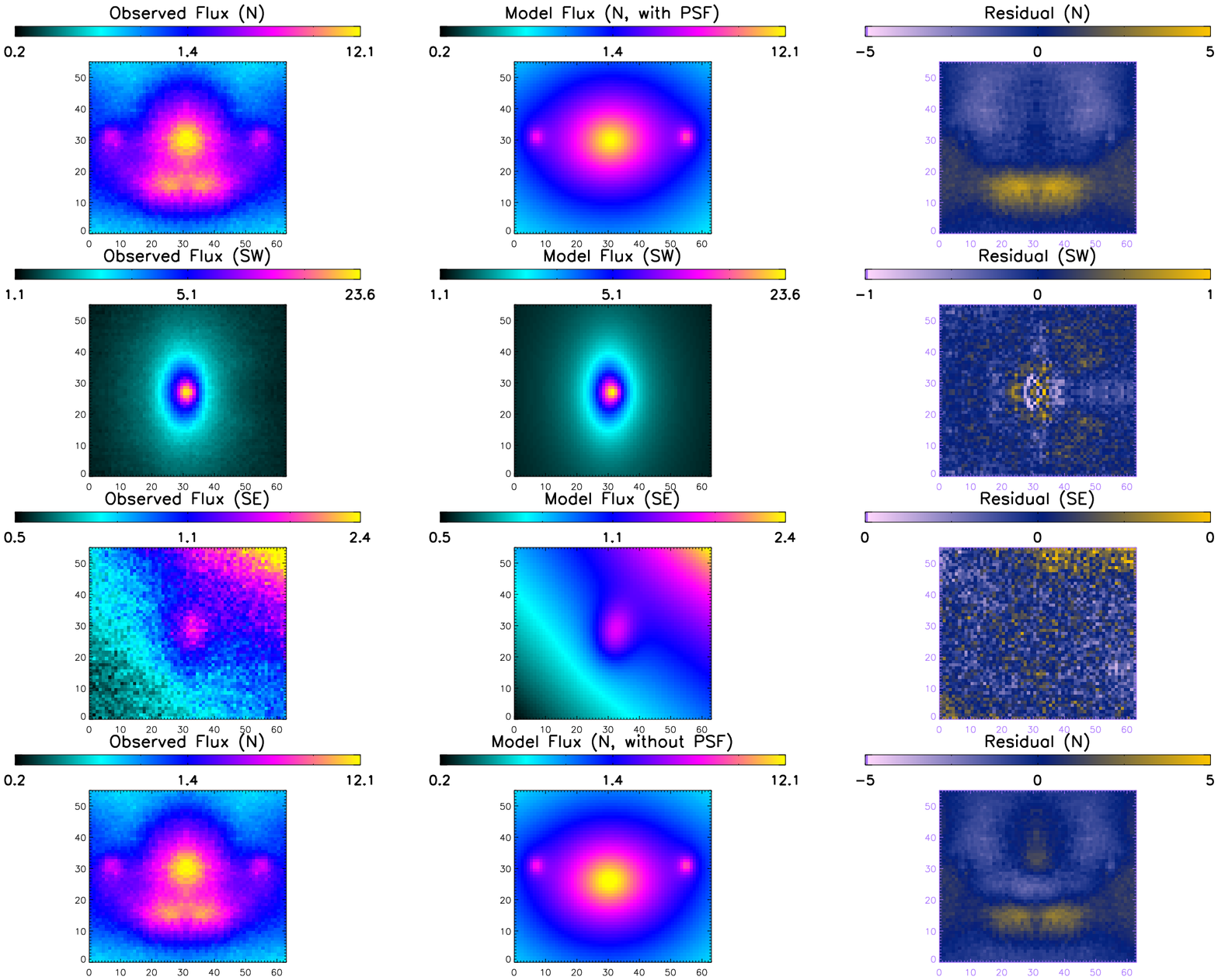}
  \caption{Example of fitting results of GALFIT: here, the N (top
    row), SW (second row), SE (third row), and N (without central PSF
    in model; bottom row) components in the NIRC2 $H$-band image have been
    fitted with a S\'ersic, S\'ersic, Gaussian, and S\'ersic profile,
    respectively. The leftmost column shows the input data that were
    fed into GALFIT; the middle column demonstrates the best-fit model
    parameters; the rightmost column indicates the residual maps. In the
    case of the N and SW components, we performed GALFIT analysis on a
    mirrored version of the eastern and southern halves, respectively,
    in order to avoid the dusty region between the nuclei.}  
  \label{fig:galfit}
\end{figure*}

Exhibiting a disky morphology different from that seen in the optical
bands, the N component is both extended and clumpy in the
near-infrared. 
We used GALFIT \cite[][]{Peng02,Peng10} to quantitatively characterize
its morphology and find that the underlying structure is best fit by a  
central point source (which we call N1) plus a S\'ersic
profile of index $n = 1.63$ and effective radius $r_{\rm eff} =
0\farcs69\pm0.01$, or 520$\pm$8 pc, in the $H$-band continuum.
In order to avoid the dusty region between the nuclei, we performed
the GALFIT analysis on the eastern half of the disk only (see
\fig~\ref{fig:galfit}). The residual map shows some underlying
substructure in the disk, but otherwise a S\'ersic disk plus central
point source model (\fig~\ref{fig:galfit} top) were a better fit for
the N component than one without the central point source
(\fig~\ref{fig:galfit} bottom).  In \fig~\ref{fig:nirc2},
approximately 0$\farcs$2 to the northeast of N1 is N2,
which looks like a faint, compact clump.  It is brighter in the $K'$
band than it is in the $H$; the difference may be due to dust
extinction.  It appears 
distinct and could plausibly be the site of a stellar cluster. It
closely coincides with the N2 as designated by~\cite{Carilli00}, at the
location of a hypothesized clump of supernovae.  To the
southeast of N1 at a distance of $\sim 0\farcs15$ is N3, which is
extended and elongated in the E-W direction. It may be diffuse
emission associated with N1 given its proximity, or it may be a
separate clump.  There is a larger difference in the brightness of N1
and N3 in the $K'$-band relative to that in the $H$-band, indicating
that N1 is redder than N3.  

Lying $\sim 1\farcs0$, or 750pc, to the southwest from N1, the SW
component is also resolved at these scales with a north-south
extension ($r_{\rm eff} = 68\pm8$ pc in the $H$-band and 38$\pm$8 pc
in the $K'$-band).  The high-resolution 0$\farcs$01/pixel data further
constrain the size of this compact emission.   Detected more notably in 
the $H$-band, the north-south extended feature appears asymmetric and
elongated toward the south. The northern extension resembles a
tidally-disrupted feature, which suggests that it is material that has
not yet settled during the course of the merger.  
The SW source appears the reddest in the frame, most likely due
to dust.  This extinction obscures it entirely in the \emph{HST}
$B$-band image (\fig \ref{fig:hst_overview}). 

Two distinct sources can be seen in the southeast region relative to the
N disk. The southernmost one is compact and appears to be a star
cluster.  The one denoted with a cross (see \fig \ref{fig:nirc2}) is more
diffuse and extended; it marks the site of the SE source seen in radio
images \cite[][]{Carilli00, Bondi05}. Its spatial extent is larger
than the compact source to the south from which the PSF is
confirmed. This suggests that the SE component may not simply be a
stellar cluster as was suggested by the NICMOS 
images~\cite[][]{Scoville00}.  Given its proximity to the N component and
the morphology of the low surface-brightness feature between it and
the N component, the SE source may be associated with the N component.

    \subsection{$K$ Broadband Integrated OSIRIS Spectra}
    
\begin{figure*}
  \centering
    \includegraphics[width=0.8\textwidth]{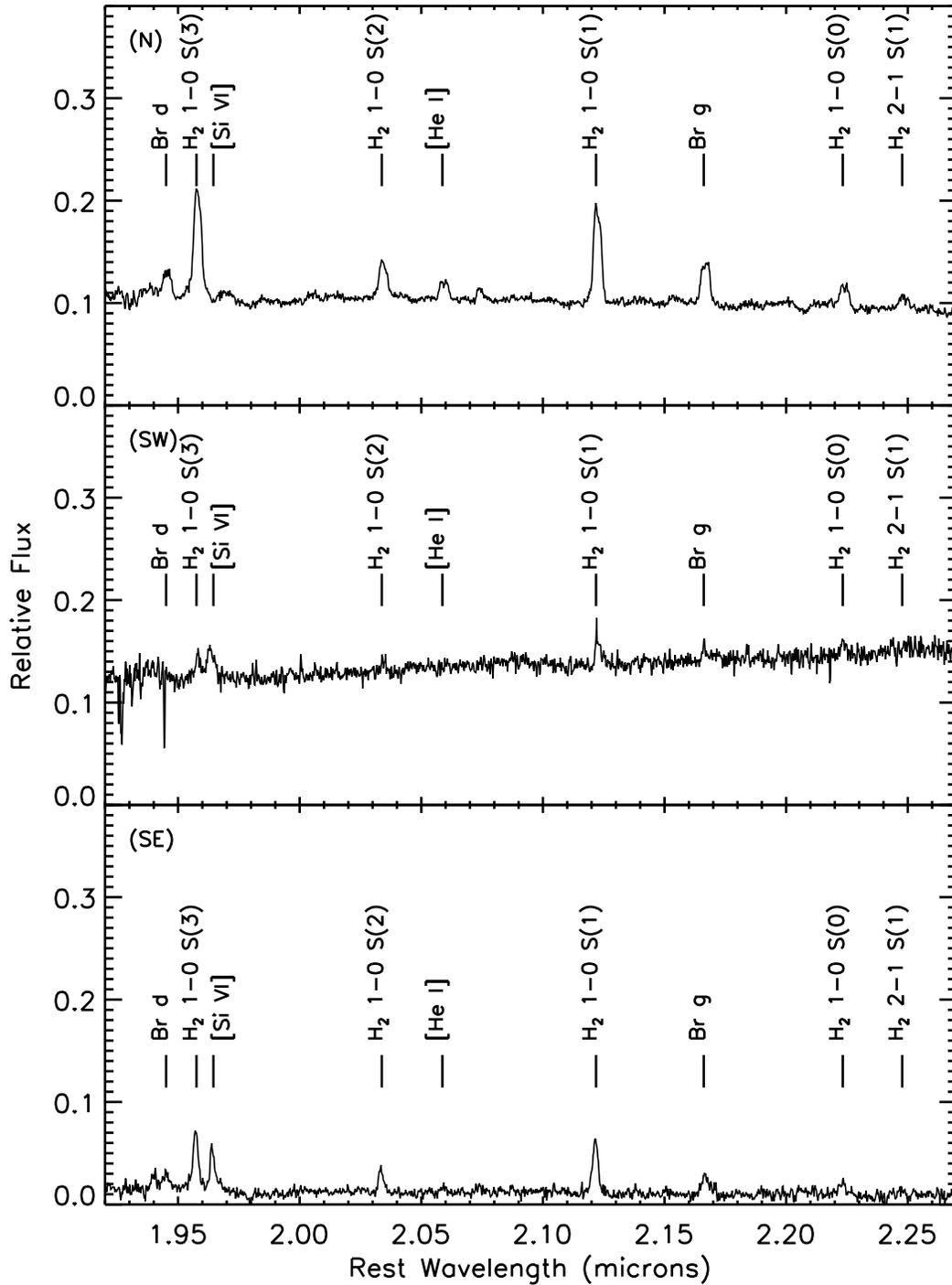}
  \caption{
    OSIRIS integrated spectra of the N component (top), the SW
    component (middle), and the SE component (bottom) from 50-minutes
    of on-source exposure, taken in $K$-band with the $0\farcs1$/pixel
    lenslet.  Lines of interest are labeled.  The
    telluric-corrected continuum in the SW still shows a steeper
    gradient than in the N and SE, indicating the presence of hot
    dust.  The SE spectrum shows comparatively little continuum
    emission but clear evidence of line emission.  Most 
    emission lines in the SW are relatively weak. On the other hand,
    \sivi~appears strongly in the SW and the SE but not in the N component.
  }  
  \label{fig:kcb}
\end{figure*}

In the $K$ broadband spectra ($0\farcs1$-per-pixel resolution), only
two spatially-segregated
components appear in the continuum of Kcb: the N extended emission and
the SW compact source. The SE diffuse clump is very weak in the
integrated image. \fig \ref{fig:kcb} shows the integrated spectra of
the N, SW, and SE components, respectively. These
spectra exhibit very different features.  The SW spectrum is dominated
by the continuum and has strong [Si VI] 1.964 $\mu$m and
\molhy~emission; other emission lines are present but weak.  The N
spectrum features multiple extremely strong \molhy~transitions along
with \brg~and \hei. The SE component is weak in the continuum but
exhibits emission in the \molhy, \sivi, and Brackett lines as well. 

Line parameters of the various regions such as the 
effective radii for the fitted S\'ersic component or FWHM of the
fitted Gaussian component, axis ratios, integrated flux ratios, mean
velocities, and velocity dispersions are listed in \tbl 
\ref{tbl:param}. The integral-field capability of OSIRIS allows the
examination of the spatial distributions and kinematics of these
emission lines individually; we present these results in the following
section. 

 \begin{deluxetable*}{rlcccccc}
    \centering
    \tabletypesize{\tiny}
    \tablewidth{0pt}
    \tablecolumns{8}
    \tablecaption{Parameters of the Different Regions}
    \tablehead{   
      \colhead{Region} &
      \colhead{Line} &
      \colhead{$r_{\rm eff}$} &
      \colhead{$r_{\rm eff}$} &
      \colhead{Axis Ratio} &
      \colhead{Flux Ratio} &
      \colhead{$\bar{v}$} &
      \colhead{$\sigma$} \\
      \colhead{} & 
      \colhead{} &
      \colhead{(arcsec)} & 
      \colhead{(pc)} & 
      \colhead{} &
      \colhead{} &
      \colhead{(km/s)} &
      \colhead{(km/s)} 
      	}
    \startdata
    N & H$_{2c}$   & 0.27$\pm$0.01 & 204$\pm$8 & 0.75$\pm$0.01 & 1.15$\pm$0.01 & 6$\pm$2 & 225$\pm$2 \\ 
      & \brg~     & 0.30$\pm$0.06 & 226$\pm$45 & 0.75$\pm$0.01 & 0.48$\pm$0.02 & 21$\pm$7 & 228$\pm$7 \\ 
      & \hei~     & 0.27$\pm$0.01 & 204$\pm$8 & 0.75$\pm$0.01 & 0.22$\pm$0.02 & 45$\pm$15 & 230$\pm$20 \\ 
      & \sivi~    & .. & .. & .. & .. & .. & .. \\
      & \feii~    & 0.40$\pm$0.04 & 302$\pm$30& 0.78$\pm$0.01 & 0.13$\pm$0.01 & 6$\pm$4 & 213$\pm$9 \\ 
      & $H$-band continuum & 0.69$\pm$0.01 & 520$\pm$8 & 0.74$\pm$0.01 & .. & .. & .. \\ 
      & $K'$-band continuum & 0.80$\pm$0.01 & 603$\pm$8 & 0.88$\pm$0.01 & .. & .. & .. \\ 
      \hline
   SW & H$_{2c}$   & .. & .. & .. & 0.29$\pm$0.03 & 50$\pm$15 & 180$\pm$15 \\
      & \brg~     & .. & .. & .. & .. & .. & .. \\
      & \hei~     & .. & .. & .. & .. & .. & .. \\
      & \sivi~    & 0.14$\pm$0.02$^\star$ & 106$\pm$15 & 0.08$\pm$0.2 & 0.37$\pm$0.06 & -290$\pm$35  & 220$\pm$50 \\ 
      & \feii~    & .. & .. & .. & 0.04$\pm$0.04 & 7$\pm$30 & 155$\pm$45 \\
      & $H$-band continuum & 0.09$\pm$0.01 & 68$\pm$8 & 0.41$\pm$0.01 & .. & .. & .. \\ 
      & $K'$-band continuum & 0.05$\pm$0.01 & 38$\pm$8 & 0.47$\pm$0.01 & .. & .. & .. \\ 
      \hline
   SE & H$_{2c}$  & 0.14$\pm$0.03$^\star$ & 106$\pm$24 & 0.56$\pm$0.05 & 0.59$\pm$0.02 & -117$\pm$4 & 173$\pm$4 \\
      & \brg~     & 0.07$\pm$0.04$^\star$ & 53$\pm$30 & 0.42$\pm$0.05 & 0.30$\pm$0.03 & 45$\pm$20 & 240$\pm$20 \\
      & \hei~     & .. & .. & .. & .. & .. & .. \\
      & \sivi~    & 0.16$\pm$0.02$^\star$ & 121$\pm$15 & 0.66$\pm$0.09 & 0.29$\pm$0.03 & -125$\pm$20 & 150$\pm$15 \\
      & \feii~    & 0.06$\pm$0.01$^\star$ & 45$\pm$8 & 0.23$\pm$0.05 & 0.06$\pm$0.04 & -10$\pm$40 & 205$\pm$60 \\
      & $H$-band continuum & 0.06$\pm$0.01$^\star$ & 45$\pm$8 & 0.61$\pm$0.04 & .. & .. & .. \\ 
      & $K'$-band continuum & 0.13$\pm$0.01$^\star$ & 98$\pm$8 & 0.51$\pm$0.01 & .. & .. & .. \\ 
      \hline
   Bridge & H$_{2c}$  & 0.35$\pm$0.03$^\star$ & 264$\pm$23 & 0.68$\pm$0.04 & 0.73$\pm$0.02 & -125$\pm$4 & 175$\pm$5 \\
      & \brg~     & 0.38$\pm$0.14$^\star$ & 287$\pm$106 & 0.16$\pm$0.06 & .. & .. & .. \\
      & \hei~     & .. & .. & .. & .. & .. & .. \\
      & \sivi~    & 0.13$\pm$0.03$^\star$ & 98$\pm$23 & 0.17$\pm$0.06 & .. & .. & .. \\
      & \feii~    & .. & .. & .. & 0.10$\pm$0.01 & -110$\pm$15 & 240$\pm$20 \\
      & $H$-band continuum & .. & .. & .. & .. & .. & .. \\ 
      & $K'$-band continuum & .. & .. & .. & .. & .. & .. 
    \enddata
    \label{tbl:param}
    \tablenotetext{1}{$r_{\rm eff}$ is the effective radius for the
      fitted S\'ersic component; in the $^\star$ cases, this is the
      $\frac{{\rm FWHM}}{2}$
      of the fitted Gaussian component.}
    \tablenotetext{2}{.. indicates weak or no detection.}
  \end{deluxetable*}

       \subsubsection{\molhy}
Five vibrational transitions of molecular hydrogen lie within the Kcb
coverage of OSIRIS: $1-0$ S(3) [$\lambda_{\rm rest}$ = 1.9576$\mu$m],
$1-0$ S(2) [$\lambda_{\rm rest}$ = 2.0338$\mu$m], $1-0$ S(1)
[$\lambda_{\rm rest}$ = 2.1218$\mu$m], $1-0$ S(0) [$\lambda_{\rm
  rest}$ = 2.2235$\mu$m], and $2-1$ S(1) [$\lambda_{\rm rest}$ =
2.2477$\mu$m] (hereafter H$_{2a}$, H$_{2b}$, H$_{2c}$, H$_{2d}$, and H$_{2e}$,
respectively).  These vibrationally-excited emission lines trace the
warm molecular gas, and result from a mixture of thermal and
nonthermal components, where the excitation mechanisms may 
involve shock heating, UV photoioniziation by O and B stars, or X-ray
ionization~\cite[][]{Mouri94}.  The intensity ratios among these
lines can be used to distinguish between the thermal and non-thermal
excitation mechanisms due to the difference in their efficiencies in
populating the various vibrational levels.  

\begin{figure*}
    \centering
    \includegraphics[width=0.8\textwidth,angle=90]{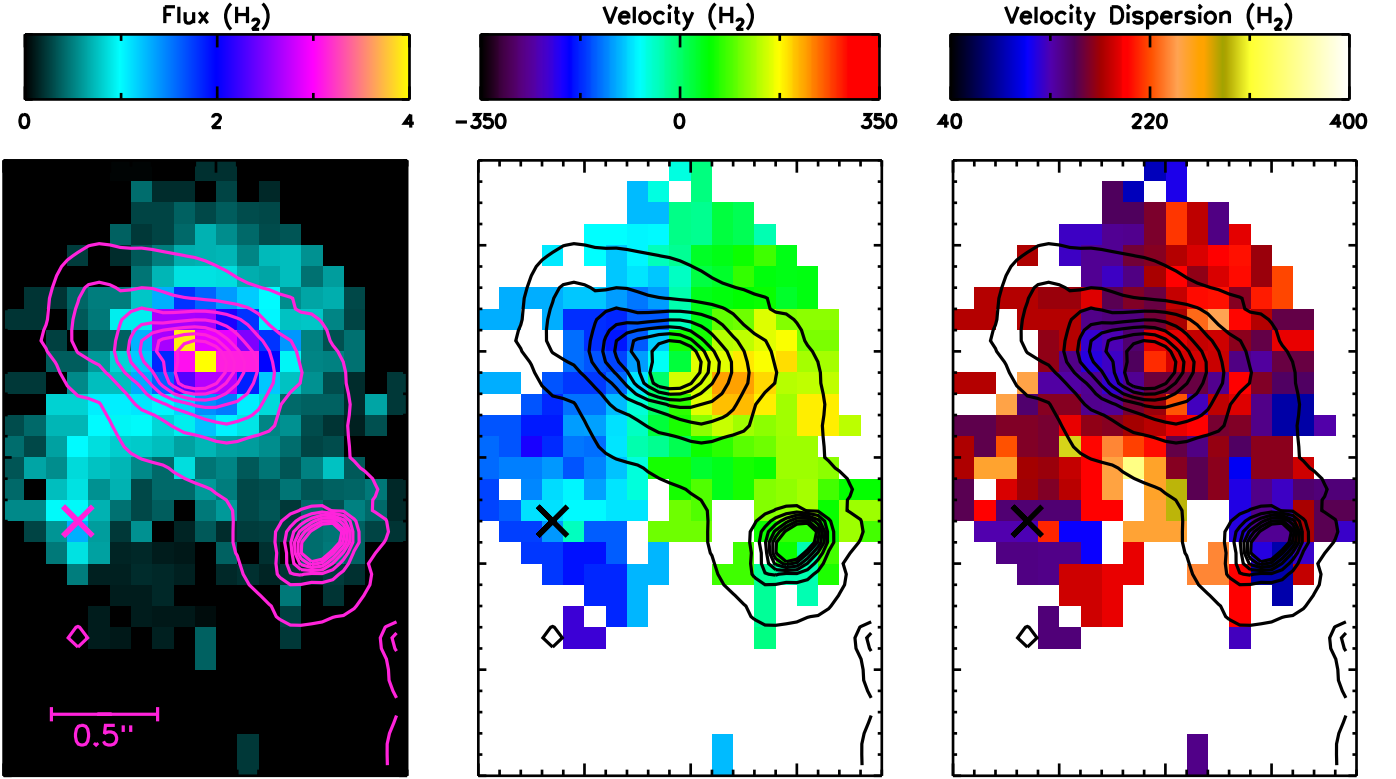}
    \caption{\emph{(Left to right)} The moments 0, 1, and 2 maps
      corresponding to the unbinned flux, binned velocity, and
      binned velocity dispersion maps of H$_{2c}$ in the central
      region of Mrk 273. Pixels with low signal-to-noise ratio have
      been removed from the kinematics maps.  Contour levels represent
      the continuum flux, with cross marking the SE region. 
      Bar denotes 0$\farcs$5; N points up.  A bridge in flux connects
      the N and SE components.  The bridge shows negative velocities
      and increased velocity dispersions.  The velocity map also
      reveals strong rotation in the N component; here, velocity
      dispersion increases in conical regions along the minor axis,
      possibly revealing biconical collimated outflows from the center
      of the disk.}    
    \label{fig:binned_h2}
  \end{figure*}

Shock-excited \molhy~emission has been observed in many galaxies; a
notable object among these observations is the prototypical late-stage
merger NGC 6240 with dual AGN at $L_{\rm IR} = 10^{11.93} L_\odot$
, for which the thermal \molhy~temperature was determined to be 2000
K~\cite[][]{Mouri94}.  \cite{Draine90} further proposed that most of
the \molhy~line emission in NGC 6240 is attributed to X-ray
irradiation from either supernova explosions or high-velocity shock
waves associated with the merger. Like NGC 6240, Mrk 273 displays
prominent emission in H$_{2c}$. \fig \ref{fig:binned_h2} 
shows the unbinned flux, Voronoi-binned velocity and velocity
dispersion maps of H$_{2c}$ along with the distribution of the $K$
band continuum in contours. The H$_{2c}$ flux is extended in the N component. 
In addition, the flux bridges toward the SE component, which is the
site of radio continuum emission ($\sim 1\arcsec$ southeast of the N
peak).

A velocity gradient typical of an inclined rotating disk can be seen in
the velocity map of H$_{2c}$ (middle panel in \fig \ref{fig:binned_h2}).
The range of the velocity is [-206, 220] km s$^{-1}$; the rotational
axis of the rotating disk has a PA of 330$^\circ$.
The velocity map spans a larger region beyond the extent of the N
disk. Of particular note are the negative velocities extending toward
the SE component along the bridge in the emission. Along the major
axis of the disk, we measure a velocity dispersion of $\sim$ 140 km
s$^{-1}$; along the minor axis and toward the SE, this increases to
$\sim$ 200 km s$^{-1}$ (right panel in \fig \ref{fig:binned_h2}). 
This lends evidence to the presence of biconical collimated turbulent
outflows from the center of the N source; the significance of this
outflow is further discussed in \S \ref{Discussion}. 

    \subsubsection{Excitation Mechanisms of \molhy}
Multiple \molhy~transitions are prominent in the Kcb data,
particularly within the N disk. Using equations from \cite{Reunanen02}
and \cite{Rodriguez04,Rodriguez05} that make use 
of the ratios among the various \molhy~transitions (\eg
H$_{2e}$/H$_{2c}$ and H$_{2b}$/H$_{2d}$), we can gain insight into the
temperature conditions within Mrk 273's nuclei.  The rotational
temperature does not vary significantly across the nuclear regions. 
Integrating over the regions defined by the different components,
we derive for the N disk a vibrational temperature $T_{\rm vib}
\simeq 2590\pm80$K and a rotational temperature $T_{\rm rot} \simeq
1605\pm115$K. The measured $T_{\rm vib}$ is consistent with that
measured for a sample of Seyferts ($T_{\rm vib} \lesssim 2600$K)
from~\cite{Reunanen02}. In addition, the fact that the values for
$T_{\rm vib}$ and $T_{\rm rot}$ differ appreciably suggests that the
thermal excitation may be due to fluorescence~\cite[][]{Rodriguez04}. 
The computed \molhy~line ratios along with derived temperatures for
the various components are  listed in \tbl \ref{tbl:h2ratios}. 

 \begin{deluxetable*}{lcccccc}
    \centering
    \tabletypesize{\scriptsize}
    \tablewidth{0pt}
    \tablecolumns{7}
    \tablecaption{\molhy~Line Ratios}
    \tablehead{   
      \colhead{Line Ratios} &
      \colhead{N} &
      \colhead{SW} &
      \colhead{SE} &
      \colhead{Bridge} &
      \colhead{Total (Mean)} & 
      \colhead{Total (Median)} \\
      	}
    \startdata
    H$_{2a}$/H$_{2c}$ & 1.19$\pm$0.02 & 0.31$\pm$0.07 & 0.55$\pm$0.03 & 0.75$\pm$0.04 & 0.70 & 0.75 \\
    H$_{2b}$/H$_{2c}$ & 0.45$\pm$0.01 & 0.11$\pm$0.05 & 0.27$\pm$0.03 & 0.26$\pm$0.03 & 0.27 &  0.27 \\
    H$_{2d}$/H$_{2c}$ & 0.29$\pm$0.01 & 0.15$\pm$0.05 & 0.17$\pm$0.02 & 0.20$\pm$0.03 & 0.20 & 0.20 \\
    H$_{2e}$/H$_{2c}$ & 0.18$\pm$0.45 & 0.04$\pm$0.12 & ... & ... & 0.11 &
    0.18 \\
    $T_{\rm vib}$ (K) & 2590$\pm$80  & 2535$\pm$740 & 2240$\pm$230 &
    1925$\pm$305 & 2325 & 2535 \\
    $T_{\rm rot}$ (K) & 1605$\pm$115 & 740$\pm$150 & 1645$\pm$355 &
    1230$\pm$205 & 1305 & 1605
    \enddata
    \label{tbl:h2ratios}
  \end{deluxetable*}

\molhy~gas in galaxies can be excited by a number
of mechanisms in the interstellar medium: UV fluorescence,
shocks, and X-ray heating~\cite[][and others]{Shull82}.  A diagnostic
involving the ratios of different \molhy~transitions can be used to
distinguish among the various means of exciting the
gas~\cite[][]{Rodriguez04}.  The H$_{2e}$/H$_{2c}$ and
H$_{2a}$/H$_{2c}$ ratios in these nuclear regions are consistent with
slow shocks ($v_s \sim 10-14$ km s$^{-1}$) being driven into very dense
gas~\cite[$n = 10^5~{\rm cm}^{-3}$;][]{Shull78}.  

       \subsubsection{Atomic Hydrogen and Helium}
After the various transitions of \molhy, the \brg~line [$\lambda_{\rm
rest} = 2.166\mu$m] and the \hei~line [$\lambda_{\rm rest} =
2.059\mu$m] are the next most prominent emission lines in the N 
spectrum. \brg~is an indicator of the ionizing radiation field and
stellar activity, and has been detected in many Seyfert
galaxies~\cite[\eg][]{Riffel06}. The ionization energy of atomic
helium is 24.6 eV; therefore, \hei~line emission is expected to
originate from regions with the most massive and youngest
stars~\cite[][]{Boker08}.  

\begin{figure*}
    \centering
    \includegraphics[width=0.8\textwidth,angle=90]{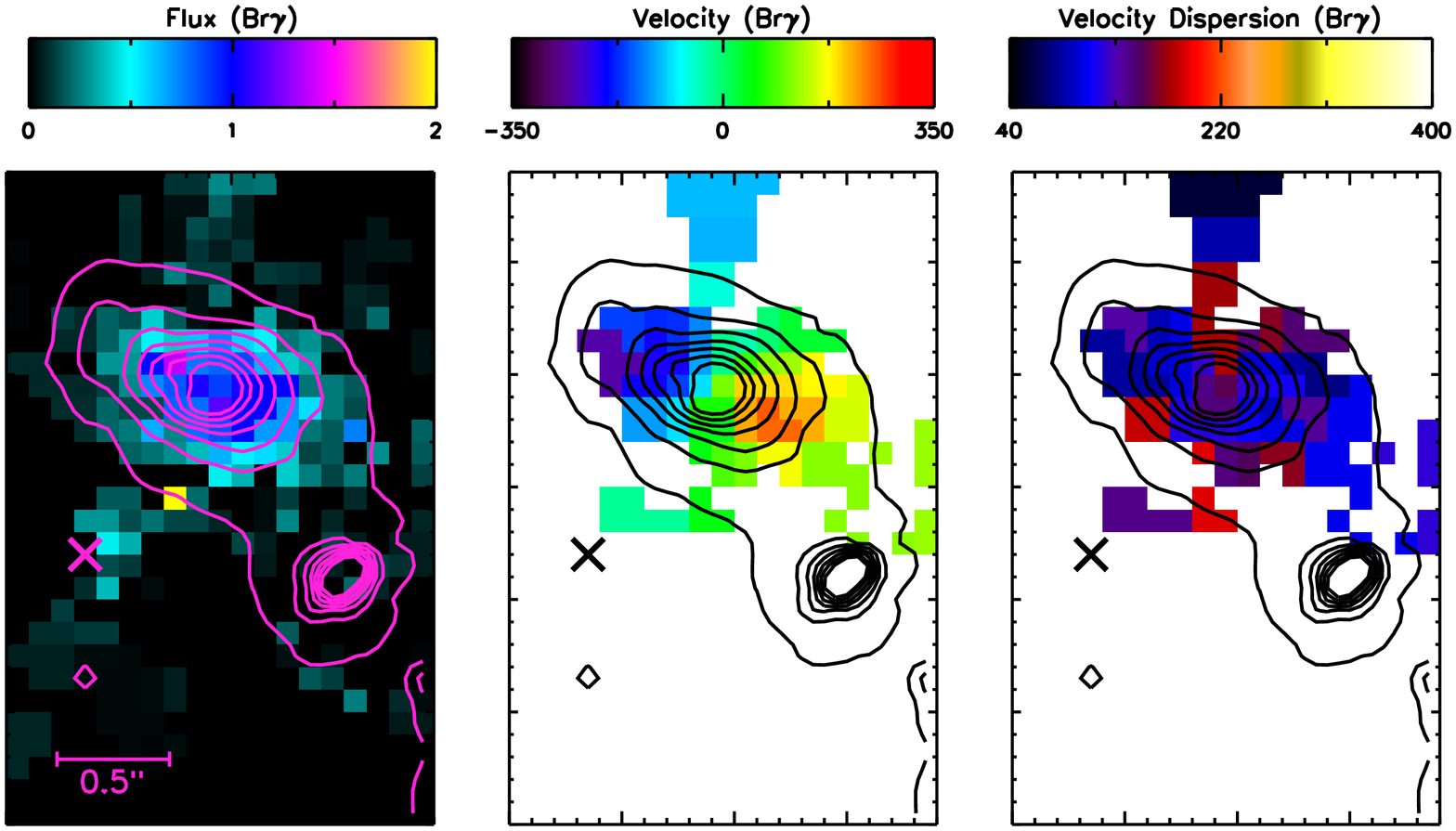}
    \caption{Same as \fig \ref{fig:binned_h2}, but for \brg~
      (N points up).  \brg~traces the same rotation in the N component
      as H$_{2c}$, but is much weaker along the bridge to the SE
      component.}    
    \label{fig:binned_brg}
  \end{figure*}

\begin{figure*}
    \centering
    \includegraphics[width=0.8\textwidth,angle=90]{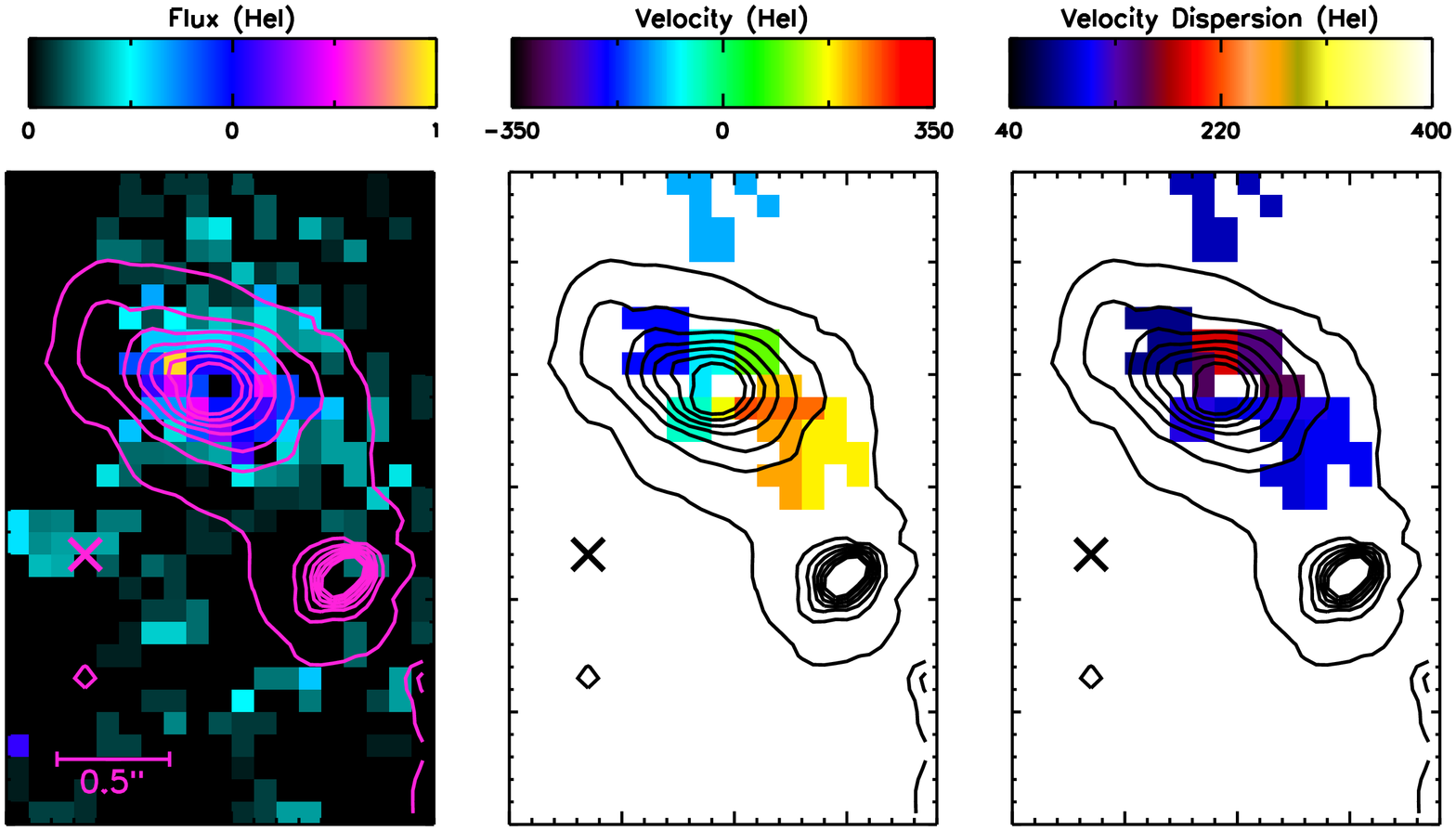}
    \caption{Same as \fig \ref{fig:binned_h2}, but for \hei~
      (N points up).
      \hei~traces the same rotation in the N component as H$_{2c}$ and
      \brg, but is much  weaker along the bridge to the SE component.  As with \brg,
      \hei~flux does not peak at the center of the N component,
      perhaps due to extinction or a change in the ionizing radiation.}  
    \label{fig:binned_hei}
  \end{figure*}

\fig \ref{fig:binned_brg} and \fig \ref{fig:binned_hei} show the flux,
velocity, and velocity dispersion of the \brg~and \hei~lines,
respectively.  
The \brg~flux, and similarly the \hei~flux, is detected in the N disk
and appears clumpy around the center; this may be due to clumpy star
formation in the circumnuclear regions. The depression in flux of
these atomic gases at the nucleus may suggest heavy obscuration by
dust or ionization of the gas by a strong source like an AGN. The bulk of
the \brg~gas coincides with the N continuum emission.  There is very
little \brg~emission at the SW component, but there is \brg~emission
extending toward the SE source, albeit much weaker than in the \molhy~lines.
There is also faint \hei~emission in the vicinity of the SE source.  

\brg~and \hei~also display similar kinematics.  The velocity
structure of a strongly rotating disk similar to that of 
\molhy~is found for \brg~and \hei. The centers of the rotating gas
disks coincide with the peak of the continuum flux. 
The velocity dispersion at the disk is $\sim$100 km s$^{-1}$ but increases 
to $\sim$ 170 km s$^{-1}$ at the center.  We see no evidence
of a broad line region.

       \subsubsection{\sivi}
       \label{Other}

\begin{figure*}
    \centering
    \includegraphics[width=0.8\textwidth,angle=90]{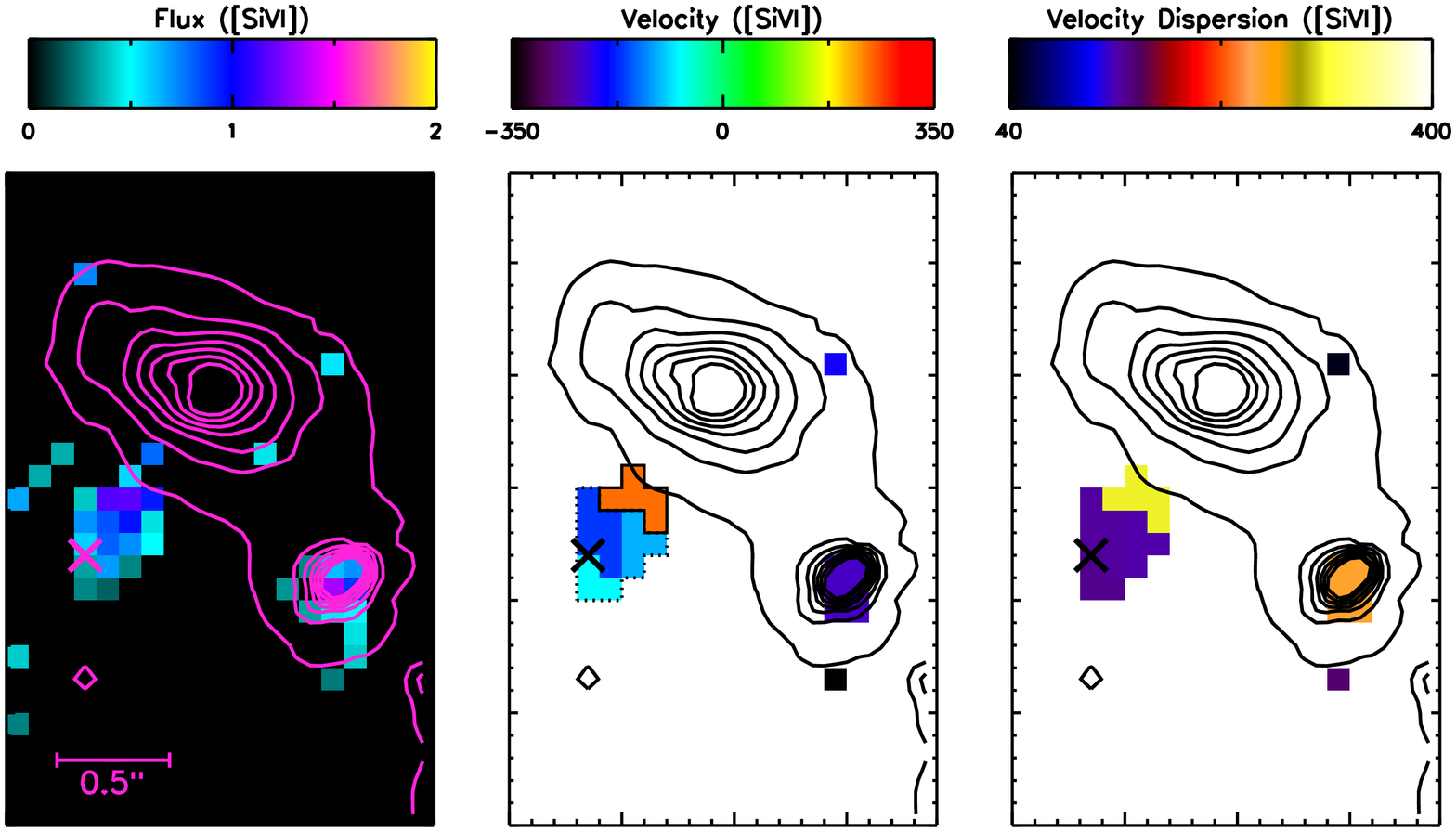}
    \caption{Same as \fig \ref{fig:binned_h2}, but for \sivi~
      (N points up).
      \sivi~is undetected in the N component, but appears strongly in
      the SE and SW components.  In the SW, \sivi~is broad, but
      see \fig\ref{fig:twocomp} for a detailed analysis of the SE
      \sivi~spectrum, which shows two kinematic components.  The two
      kinematically and spatially segregated regions in
      \fig\ref{fig:twocomp} are marked in the center panel: the
      northwest edge is outlined in solid black, the remaining SE
      pixels are outlined in dotted black line.  The northwest edge 
    shows broader kinematics than the dotted region.}  
    \label{fig:binned_sivi}
  \end{figure*}

\sivi~[$\lambda_{\rm rest} = 1.964\mu$m], a near-infrared tracer of
AGN activity due to its high excitation potential~\cite[= 167 eV;][]{Rodriguez04}, 
has a very different spatial distribution than the other species, as
seen in \fig \ref{fig:binned_sivi}. A fit to the line in the SE region
with a Gaussian profile returns poor residuals. A visual inspection
prompted us to add a second velocity component which
significantly improved the fit (\fig \ref{fig:twocomp}). Two
kinematically-segregated components in this region 
are traced by \sivi: a narrow component with velocity $v_n =
-162\pm6$ km s$^{-1}$ and velocity dispersion $\sigma_n = 99\pm8$
km s$^{-1}$; and a broad component with $v_b = 65\pm35$ km s$^{-1}$
and $\sigma_b = 325\pm25$ km s$^{-1}$.

 The bulk of the \sivi~emission is found not in the N disk, likely due
 to heavy extinction, but at the bridge and the SE component. 
In the unbinned flux map (\fig \ref{fig:binned_sivi}), it extends N-S and
spans a $0\farcs189 \times 0\farcs163$ (143 pc $\times$ 123 pc)
region.  Its velocity and velocity dispersion are highest near the bridge,
though its relatively low S/N ratio limits the precision with which the
location of the emission can be determined. \sivi~is also strongly
detected at the SW component, the site of the hard X-ray
AGN~\cite[][]{Iwasawa11_mrk273}. The peak position of the \sivi~flux
roughly coincides with that of the continuum, with $v = -290\pm30$km
s$^{-1}$ and $\sigma = 222\pm50$km s$^{-1}$.

    \subsection{H Band Integrated Spectra}

\begin{figure*}
  \centering
    \includegraphics[width=0.8\textwidth]{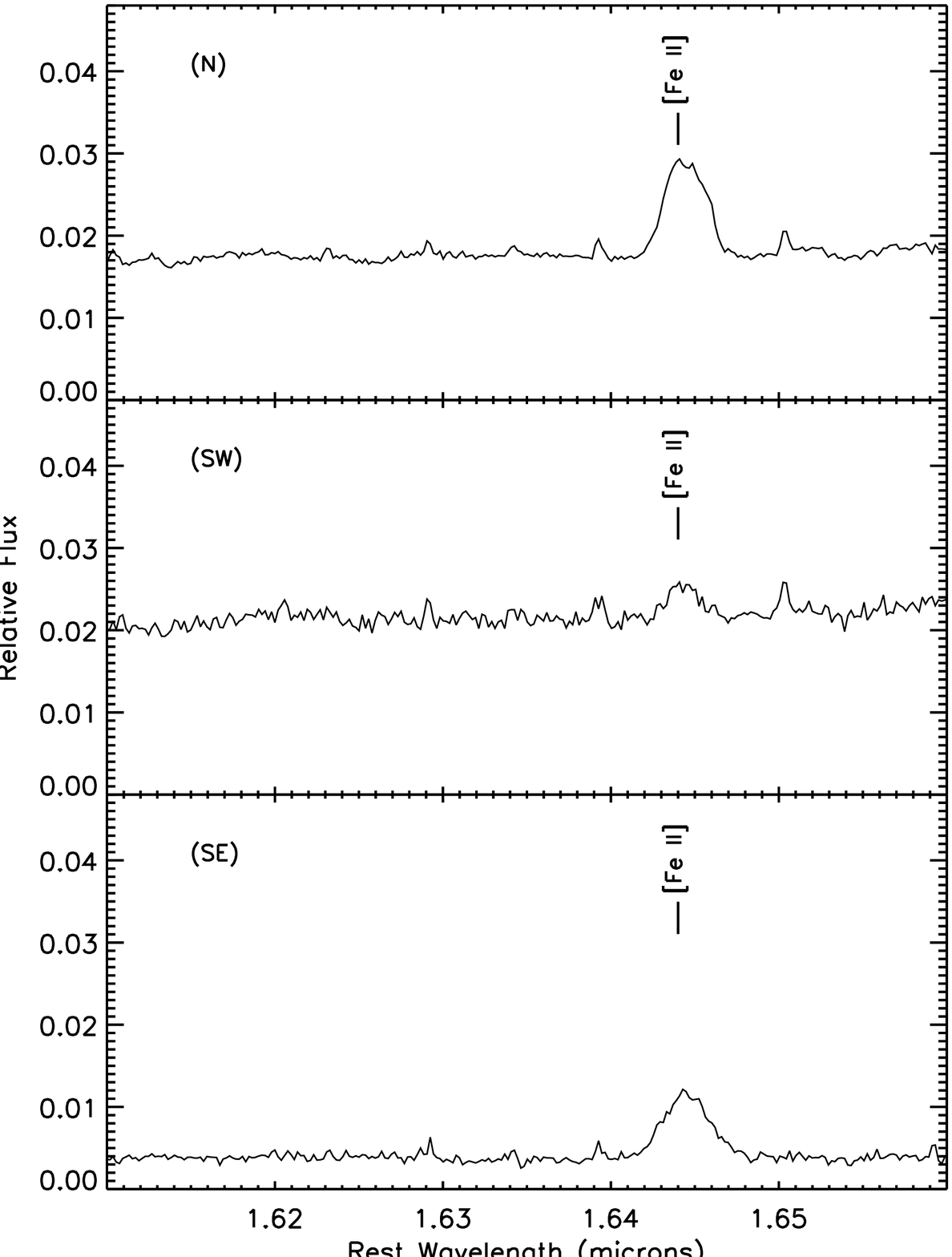}
  \caption{
    OSIRIS integrated spectra of the N (top), SW (middle), and SE
    components (bottom) from 60-minutes of on-source exposure, taken in
    $H$-band with the $0\farcs035$-scale lenslet. \feii~is much stronger
    in the N and SE components than in the SW component.
  }  
  \label{fig:hn4}
 \end{figure*}

At a scale of $0\farcs035$ per pixel, the Hn4 continuum emission maps
the resolved, clumpy N disk and compact SW component. \fig
\ref{fig:hn4} shows the integrated spectra of the N, SW, and SE
components, respectively, highlighting the \feii~line found in all
three regions. The \feii~line is typically broad ($\sigma \sim 200$ km
s$^{-1}$) and prominent in the N and SE components, and dominated by the
continuum in the SW.  The SE component is weak in the continuum.  Line
parameters of the regions for the \feii~line emission can be found in
\tbl \ref{tbl:param}.  

       \subsubsection{\feii~Emission}

\begin{figure*}
    \centering
    \includegraphics[width=0.8\textwidth,angle=90]{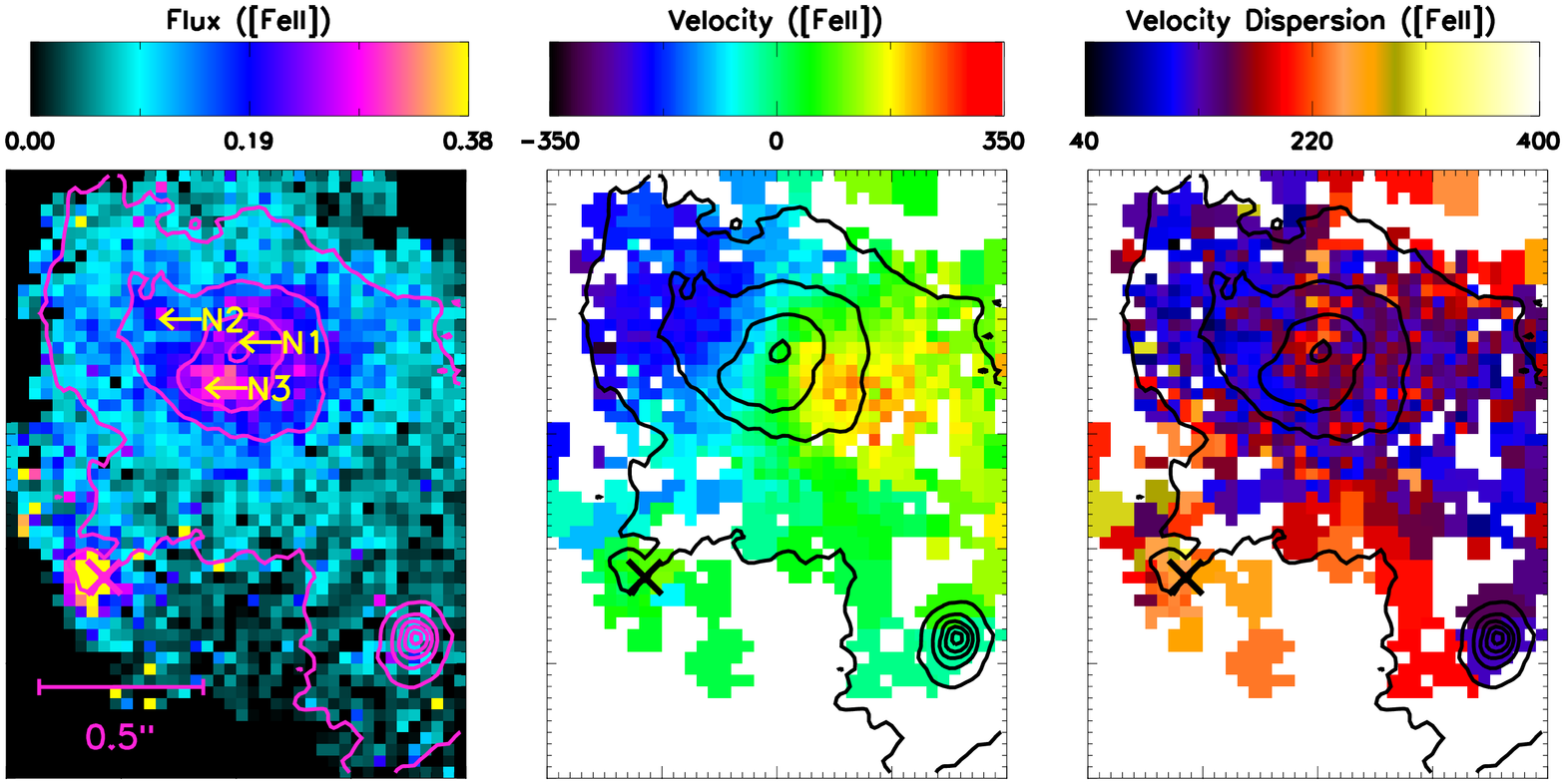}
    \caption{Same as \fig \ref{fig:binned_h2} but for \feii~
      ($\lambda_{\rm rest} = 1.644\mu m$) from the Hn4 band (N points up).
      The \feii~velocity map shows strong rotation about the N
      component, with a small increase in velocity dispersion at the
      center.  \feii~flux is clumpy in the N component, resolving N1,
      N2, and N3 due to high resolution as in the NIRC2 images
      (\fig\ref{fig:nirc2}). However, the fact that N1 is weaker than
      N2 and N3 could be due to extinction or to differing ionizing
      radiation fields.  The strongest and broadest \feii~appears in
      the SE component.}   
    \label{fig:binned_feii}
  \end{figure*}

In the Hn4 band, the \feii~line at [$\lambda_{\rm rest} = 1.644 \mu$m]
shows strong emission in the center of Mrk 273.  Because Fe is strongly
depleted in the interstellar medium, gas-phase \feii~emission~\cite[with
ionization potential of 16.2 eV;][]{Mouri93} in galaxies has been used
as a sensitive indicator of shocks~\cite[][]{Graham87}. The potential
origins for the 
\feii~emission include regions of partially ionized gas within
supernova remnants in starburst nuclei and within narrow-line regions
in AGN~\cite[][]{Mouri93}.  The contribution from photodissociation
regions near O and B stars is negligible.  Extracting
slices associated with the \feii~line from the Hn4 data cube, we
present the flux, velocity, and velocity dispersion maps in \fig
\ref{fig:binned_feii} with continuum contours for comparison.  Like
the continuum, the line data are resolved into multiple clumps in
the N disk, but the emission is relatively weak at the N1 site,
dominant at N2, and detected at N3. The differences in the
line-to-continuum flux ratios between N1 and the other clumps (N1:
0.016, N2: 0.022, N3: 0.022) may suggest
that they have different origins~(see \S \ref{Discussion}). The
\feii~line flux is negligible in the SW nucleus but very bright
in the SE, with a slightly N-S elongated morphology.  The N disk
contains more total integrated \feii~flux, but the SE peak dominates
the surface brightness in the entire inner kpc region.  
The velocity structure of the \feii~line also indicates strong rotation 
($\pm 240$ km s$^{-1}$ centered on the systemic velocity), with high
velocity dispersion near the center of the N disk and the strongest
velocity dispersion at the SE component.  

  \section{Discussion}
  \label{Discussion}
  The mechanisms that power the enormous luminosities in Mrk
  273 are likely to be associated with the three nuclear components
  situated within the dusty core.  Here we discuss the nature 
  of these regions. 

    \subsection{The Northern Disk}
The nature of the N component has been controversial due to different
interpretations of the multiwavelength data: clustered radio
supernovae~\cite[][]{Smith98}, Seyfert 2 AGN~\cite[][]{Xia02}, and a
starburst with the possibility of an additional Compton-thick
AGN from excess 6$-$7 keV emission that might be associated
with a high equivalent-width Fe K emission~\cite[][]{Iwasawa11_mrk273}.
The Fe K line as indicative of a heavily obscured AGN is also present in a
dual-AGN LIRG merger Mrk 266, where the LINER nucleus exhibits enhanced
shock-excited \molhy~emission and diminished coronal line compared to
the Seyfert nucleus~\cite[][]{Mazzarella12}.  
Here, our OSIRIS data cube provides evidence of a strongly rotating gas disk
through the distribution and kinematics of the near-infrared emission
line gas.   The strong H$_{2c}$ flux suggests localized shocks and 
fits well with the models of clustered radio
supernovae~\cite[][]{Smith98}, lending support to their hypothesis
that a starburst provides much of the power.~\cite{Carilli00} have
resolved individual compact sources that may be supernova
remnants; they suggest that the diffuse emission may be a result of
excitation from the remnants.  Our OSIRIS data resolve the individual
clumps and provide spatial information on the physical conditions of
the ionizing source in the N disk. We consider the evidence in this
section. 

    \subsubsection{Strong Rotation in Emission Lines}
The velocity maps for the emission lines (\molhy, \brg, \hei, and
\feii) detected in the N disk show structure consistent with a
strongly rotating disk.  The deprojected rotational velocity is $\sim
\pm$240 km s$^{-1}$, and its structure is indicative of Keplerian
rotation about a large mass (see \S \ref{BHmass}). We suggest that the
rotating disk in the N is the remnant nucleus of a progenitor galaxy
in this merger system, with its gas rotating about a central
supermassive black hole or a very massive central star cluster.  In
the next section we use dynamical modeling to determine the enclosed mass.

    \subsubsection{Black Hole Mass}
    \label{BHmass}

From the strong rotation inferred from the emission lines, we measure
the kinematics to understand the underlying mass profile.  Though
rotation in the N disk is seen in every emission line, we focus on the
\feii~kinematics because  these data have significantly higher
spatial resolution,  by a factor of three, compared to the other emission lines.
To measure the dynamical mass of the central source, it is crucial that we
resolve within the black hole's sphere of influence. The radius of the
sphere of influence is given by
\begin{equation}
r_{\rm BH} = \frac{GM_{\rm BH}}{\sigma^2} \quad ,
\end{equation}
where $G$ is the Gravitational constant, $M_{\rm BH}$ is the mass of
the black hole, and $\sigma$ is the velocity dispersion.  We assume a
conservative estimate for the mass of the black hole to be $M_{\rm BH}
\sim 10^9 M_\odot$, given that~\cite{Klockner04} had measured a mass
of $1.39 \pm 0.16 \times 10^{9} M_\odot$ from radio interferometry
observations of OH masers. 
The circumnuclear velocity dispersion we measure from the
\feii~line is $\sim$215 km s$^{-1}$.  Therefore, the radius of the sphere of
influence is at least $r_{\rm BH} \geq$ 95 pc.  
Given our resolution element of 0$\farcs$08, or 60 pc, there are 23
spaxels for which the enclosed mass is dominated by the central
point mass (see model description below).  These spaxels are
sufficient to provide a lever arm on black hole mass and, when
combined with the surrounding spaxels, enable a reliable dynamical
mass measurement from the \feii~line kinematics.

We follow the Keplerian disk fitting method outlined in
\cite{Medling11} to compare the measured kinematics to those of simple
models.  First, we create velocity maps using $v = \sqrt{G M /r}$ and
assuming a central point mass; we also solve for the inclination of
the disk and the position of the central point mass.  We use the
Levenberg-Marquardt least squares fitting routine MPFITFUN
\cite[][]{Markwardt09, More78} to select the parameters that best fit
the [Fe II] velocity map.  We fix the central mass position for
generalization to a more complete mass profile.  In our second set of
models, we estimate the stellar mass profile by fitting a disk to the
continuum emission.  We subtract off the central point-source from the
center of the continuum light before fitting.  We fit this using the
simple radially-varying surface density profile $\Sigma(r) =
\Sigma_{0} r^{-\gamma}$, making the mass in a ring $2 \pi  \Sigma_{0}
r^{1-\gamma} dr$.  We fit the inclination, $\gamma$, and $\Sigma_{0}$,
and find that the fitted inclination is consistent with the
inclination from the first dynamical model ($55^{\circ}$); we fit a
$\gamma$ of 1.0.  In our final set of models we adopt $v = \sqrt{G
  M(r) / r}$, with the mass profile containing a single point mass
plus a radially-varying disk mass of surface density $\Sigma(r) =
\Sigma_{0}~r^{-\gamma}$.  We fix the inclination and the exponent
$\gamma$ from our previous two models, and vary only the point mass
and the disk mass density normalization $\Sigma_{0}$.  (Note that we
allow the last to vary in order to avoid making assumptions about
mass-to-light (M/L) ratio.  On such small scales, dark matter should be
negligible, but
estimating a M/L ratio for a stellar population of unknown age with
unknown extinctions would introduce more uncertainties than allowing
it to be a free parameter.)  In each set of models, we convolve our
light-weighted model with the PSF (see discussion in \S
\ref{Observations}) before comparing it to our observations. Errors on
parameters have been estimated using a Monte Carlo approach, refitting
the black hole mass from a model with 100 iterations of statistical
errors and using the resulting distribution. 

\begin{figure*}
    \centering
    \includegraphics[width=0.8\textwidth,angle=90]{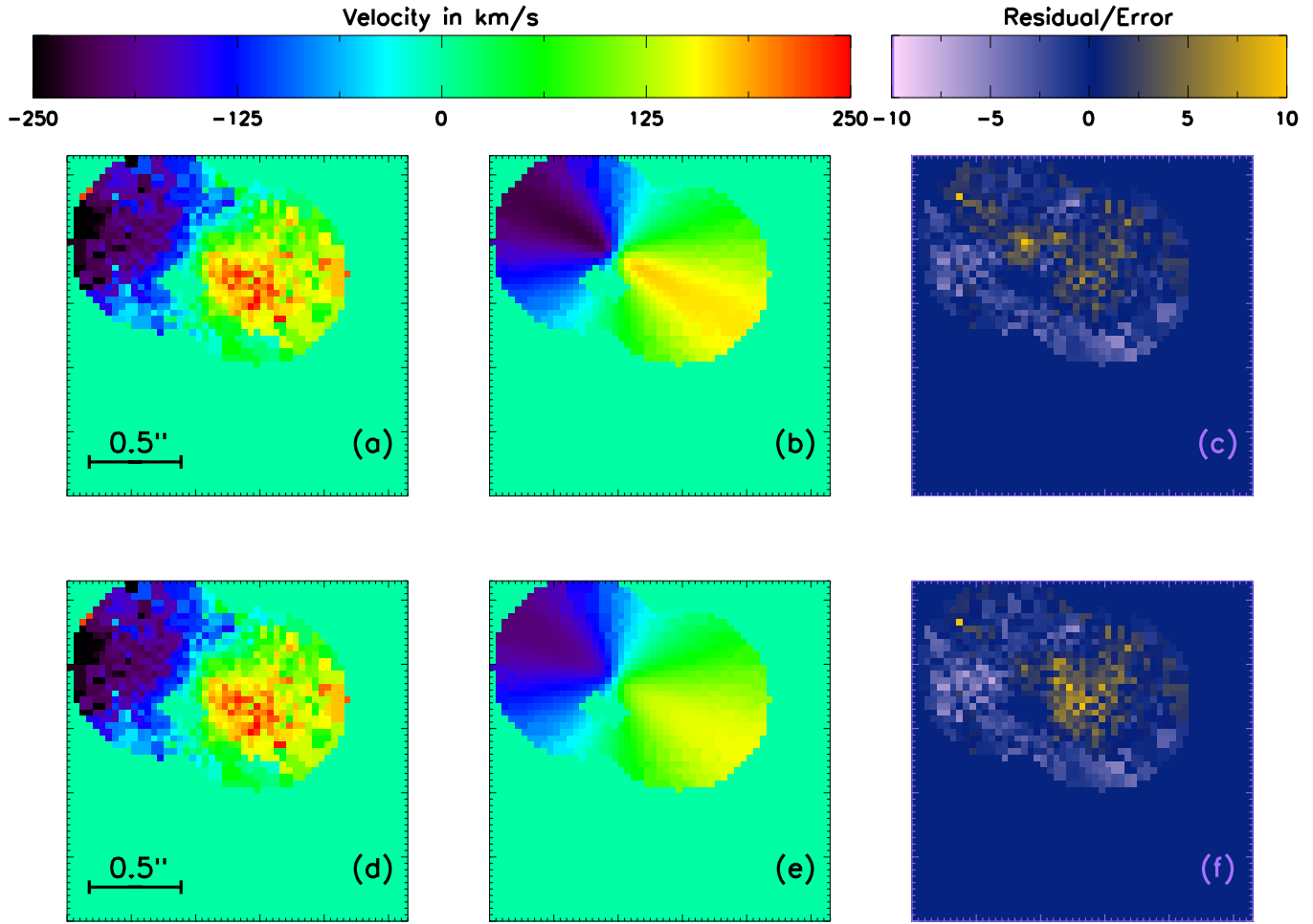}
    \caption{These panels show the kinematic modeling comparisons of
      our measured velocity map of \feii~to the velocity map of two
      rotating disk models with a mass profile containing a
      radially-varying mass component with (top row) and without
      (bottom row)
      a central point mass.  (a) data showing \feii~velocity map; (b)
      model with a point mass of $M = 1.04 \pm 0.1 \times 10^9
      M_\sun$; (c) residuals from data$-$model as normalized by the
      error; (d) same as (a); (e) model without central point mass;
      (f) residuals from data$-$model without point mass as normalized
      by the error. All of these panels are masked by a weight map
      showing only the nuclear region.  This comparison shows that the
      kinematic model incorporating a central point mass is a better
      fit to the measured velocity map. This suggests the presence of a
      black hole at the center of the N disk, and that the N is the
      remnant nucleus of a progenitor galaxy. N points up.}    
    \label{fig:BHfit}
  \end{figure*}

A comparison between the measured \feii~velocity map and the kinematic
models is found in \fig \ref{fig:BHfit}. The results illustrate that
a disk model with a central point mass turns out a better fit than one
without.  We find that the best-fit central point mass
is $ 1.04 \pm 0.1 \times 10^{9} M_{\sun}$.  This mass is consistent with 
that measured from radio interferometry of an OH maser, $1.39 \pm 0.16
\times 10^{9} M_{\sun}$~\cite[][]{Klockner04}, and is similar to that
of the southern black hole of NGC 6240 ($8.7 \times 10^{8} M_{\sun} <
M_{BH} <  2.0 \times 10^{9} M_{\sun}$), another late-stage galaxy merger
\cite[see][]{Medling11}. The mass resemblance to the NGC 6240 black
hole suggests that this central point mass may likely be a
supermassive black hole.  From the kinematics,  we cannot determine
whether Mrk 273's N disk hosts a quiescent black hole or a
Compton-thick AGN as suggested by \cite{Iwasawa11_mrk273}.  In either
case, the presence of the disk and supermassive black hole confirms
that this is the remnant nucleus of one progenitor galaxy.  

Though we have shown from our GALFIT analysis that the infrared
continuum light profile is best fit with a model of S\'ersic disk
along with a central point source, we note the possibility that the
central mass may be an unresolved massive star cluster instead of a supermassive
black hole.  Given that massive star clusters found in NGC 6240 have 
photometric masses of $7 \times 10^5 - 4 \times 10^7
M_\odot$~\cite[][]{Pollack07} and those found in the Milky Way have
upper mass limits of $10^8 M_\odot$~\cite[][]{Murray09}, $\geq$10-30
of the most massive of these clusters must reside within Keck's
resolution limit to feature the observed kinematics. The two most massive
clusters described in~\cite{Murray09} have masses of $10^8 M_\odot$
and measured radii of $\sim$ 100 pc, suggesting that a cluster 10 times as
massive and in a much smaller volume would be improbable. Since it is
implausible that 10$^9 M_\odot$ of stars would be dynamically stable
within this compact region, we conclude that the central mass must be
a supermassive black hole.

    \subsubsection{Spatial Distribution of Line Emission}
\brg, which traces ionized gas, and \hei, which traces neutral gas,
both appear in the N disk.  These species both have moderately low
velocity dispersion in the disk, and show no sign of being associated with
a broad line region.  The coronal line \sivi~is virtually non-existent
in the N disk. However, 1.4GHz images from~\cite{Carilli00} suggested
that there may be a weak AGN present near N1.  Similarly, the 6$-$7
keV extended emission detected near around the N nucleus may be due to 
enhanced 6.4 keV Fe K line emission, lending support to the presence
of a heavily obscured AGN~\cite[][]{Iwasawa11_mrk273}.  We measure an
increase in the velocity dispersion for \molhy~along the minor axis
and toward the SE, which gives direct evidence for the presence of
biconical collimated molecular outflows from the N source.  

There is a deficiency of \brg~and \hei~flux near N1 compared with the
rest of the N disk. This could be due to ionization caused
either by an AGN or by supernovae in the center of the disk. However,
we find other putative supernovae sites that do not show a deficiency in
\brg, so factors such as extinction may contribute toward this
phenomenon near N1. Our Kcb
line data are unable to fully resolve the individual clumps in the N disk, 
but the \feii~line and NIRC2 images resolve them at the
0$\farcs$035 and 0$\farcs$01 scales, respectively.  In particular, N1
is bright in the $K'$- and $H$-band flux but relatively weak in \feii,
whereas the faint clump N2 and extended clump N3 have a higher
\feii-to-continuum ratio.  N3 has similar brightness as N1 in 
the $H$-band, but is weaker than N1 in the $K'$-band.  
The presence and clumpiness of \feii~may indicate regions excited by
supernova remnants~\cite[][]{Alonso97}. However, since N1
is weak in \feii~and redder in color than N2 and N3, its ionizing
source may be similar to the AGN in the SW component.  Given that the
supermassive black hole measured dynamically in \S\ref{BHmass} lies
inside N1, we note the possibility that it may be an obscured AGN.

    \subsection{The Southwestern Nucleus}
The SW component is the location of the hard X-ray source from an
updated \emph{Chandra} map~\cite[][]{Iwasawa11_mrk273}.  It is red and
unresolved in NICMOS imaging~\cite[][]{Scoville00}, and shows 8.4GHz flux
consistent with a radio-quiet AGN~\cite[][]{Condon91}. 
In our OSIRIS data, this SW region is compact and dominated by the
continuum emission.  It shows strong \sivi~and moderate
\molhy~emission, but is weak in \brg~flux.  
At the 0$\farcs$01 scale of the NIRC2 data, the SW source is resolved
and exhibits a N-S extended, asymmetric low surface-brightness structure.  
The presence of hot dust as inferred from the rising continuum slope
agrees with the redness seen at this source in the NICMOS $JHK$-band
images~\cite[][]{Scoville00} as well as with the \emph{Spitzer}-IRS
results~\cite[][]{Armus07}, consistent with the presence of an AGN. 

The SW nucleus features low detection from \molhy, \brg, and
\hei, all potentially partially destroyed by a strong ionizing
source. \sivi~emission is detected in this region, but is not resolved. 
It usually indicates the presence of the hard ionizing 
spectrum of an AGN, consistent with the \emph{Chandra}
results that an AGN resides in this source~\cite[][]{Iwasawa11_mrk273}.

    \subsection{The Southeastern Component and its Bridge}

Though the SE component has been observed in several bands, its nature
has remained unclear and controversial. 
From \emph{HST} ACS and NICMOS images, the SE component resembles a
star cluster~\cite[][]{Scoville00} and is identified as a candidate blue star
cluster~\cite[CSC1;][]{Iwasawa11_mrk273}.  Radio images show
prominent emission and extended morphology consistent with AGN jets, but
lack the compact core indicative of an AGN.  The radio data show 
a steep spectrum ($\alpha = 1.4 \pm 0.2$), which could point to a starburst 
origin~\cite[][]{Carilli00,Bondi05}, though the extended nature ($\sim230$ pc)
suggests that it is larger than a single star cluster.  

\begin{figure*}
    \centering
    \includegraphics[width=0.8\textwidth,angle=90]{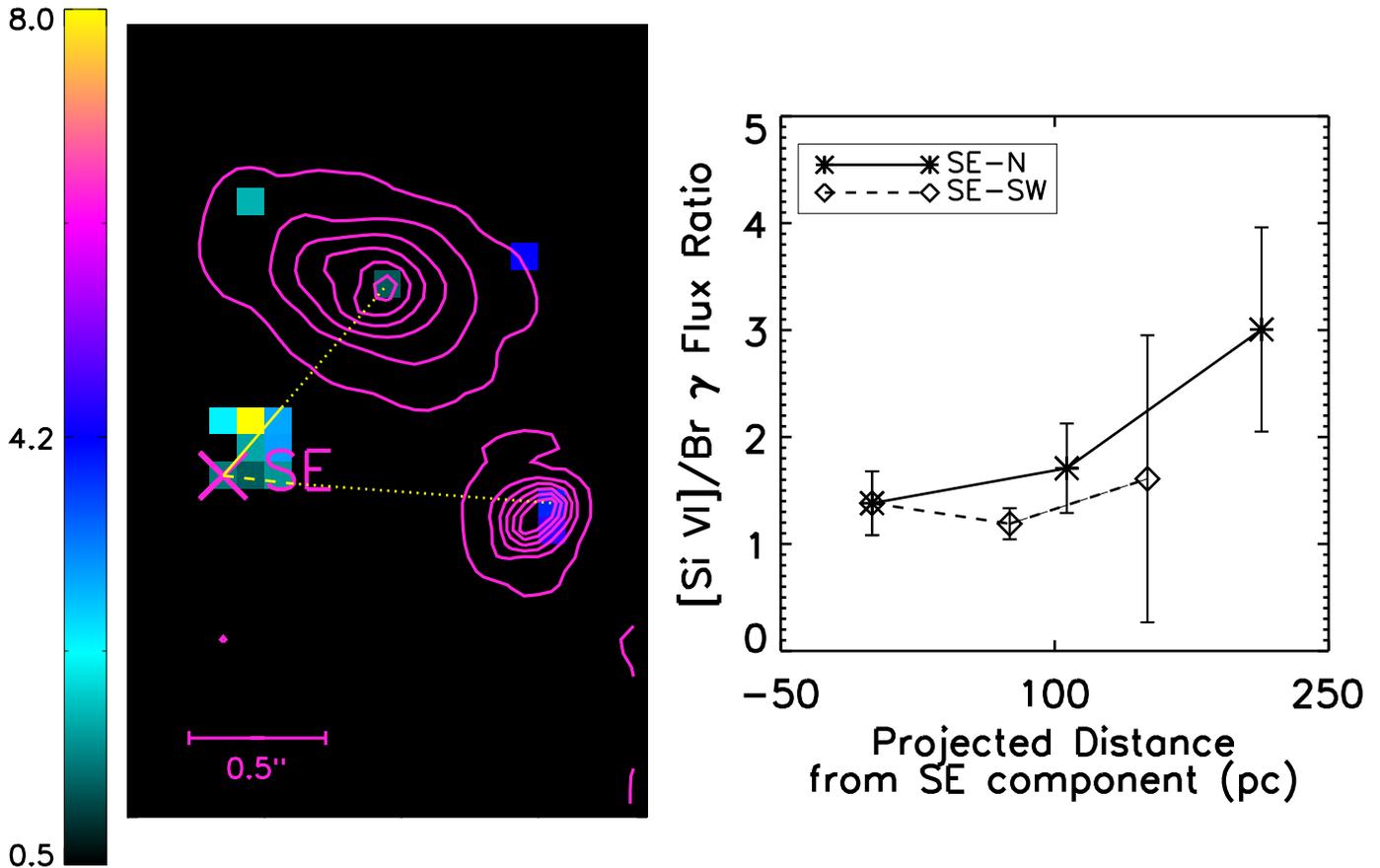}
    \caption{(Left) \sivi/\brg~ratio map with contours showing K band
      continuum emission.  The ratio appears strongest near the
      northwest boundary between the SE component and its bridge and
      decreases toward the SE component. This gradient (solid+dotted yellow
      line) suggests that the origin of the ionization mechanism may
      be external rather than local. For comparison, a dashed+dotted yellow
      line is drawn in the direction pointing to the SW nucleus. Bar
      denotes 0$\farcs$5; N points up. (Right) The radial profile of
      the \sivi/\brg~ratio plotted as a function of projected distance
    from the SE component.  The solid and dashed lines correspond to
    those in the left panel. }  
    \label{fig:sivibrgmap}
  \end{figure*}

Our spectral decomposition of the SE component in the near-infrared adds new
information. The SE component appears weak and diffuse in the NIRC2
$H$- and $K'$-band continuum images, but does appear clearly in
several emission lines: \molhy~(\fig\ref{fig:binned_h2}),
\sivi~(\fig\ref{fig:binned_sivi}), and
\feii~(\fig\ref{fig:binned_feii}) in our IFU data.
In particular, we note the strong \sivi~emission; as \sivi~has the highest 
ionization potential of the observed lines (167 eV), its presence
gives a strong clue to the origin of the SE component.  The use of
\sivi~as a tracer of the coronal line region (CLR) has been
demonstrated by~\cite{Muller-Sanchez11} in 7 nearby Seyfert
galaxies. They found that the CLRs in these low-power
Seyferts to be $\sim 80-150$ pc in radius and are associated with
outflows, providing circumstantial evidence for the existence of
extended CLRs and highly ionized gas in outflows from Seyfert
nuclei like that in Mrk 273. Here we consider two possible mechanisms
to ionize \sivi: photoionization and shock-heating.  
We first consider the \sivi/\brg~flux ratio in our models; the map
of this ratio along with the radial profiles drawn in the direction
toward the N and the SW components, respectively, are shown in
\fig\ref{fig:sivibrgmap}. We compare the \sivi~flux to \brg~flux since
the ratio may give clues to the ionization mechanisms involved. We
find that, along the northwest boundary between the SE component and
its bridge, the \sivi/\brg~ratio reaches 7.9 near the N nucleus, and
decreases over $\sim 0\farcs3$ to 1.2 near the SE component. 
We observe a regression slope of $6.6 \times 10^{-3}$, which is
statistically significant with p-value = 0.03 by permutation.  For
comparison, we compute the regression slope for a line drawn toward
the SW nucleus to be $-3.9 \times 10^{-4}$, which is not significantly
different from zero. 
This SE-N line gradient therefore suggests that the north end of the
SE component being more highly ionized, and that the origin of the
ionization mechanism may be external; we consider below potential
sources for the ionization.   

\subsubsection{Photoionized \sivi}
\label{sec:photo}

If the \sivi~is photoionized, the hard ionizing radiation may come
from either an AGN or nearby O and B stars.  We compared our measured
\sivi/\brg~flux ratio to those produced by CLOUDY models representing
each scenario.  We used CLOUDY version 08.00, described by~\cite{Ferland98}.

First we confirmed that an AGN can plausibly produce our measured
\sivi/\brg~flux ratio.  Our incident AGN spectrum was a power-law
continuum with a spectral index of -0.2, and we assumed that the gas
is approximately solar metallicity~\cite[consistent with estimates
from][]{Huo04}.  We varied the ionization parameter of the gas $U =
Q(H^{0})/(4 \pi r^{2} n_{H} c)$, where $Q(H^{0})$ is the flux of
ionizing photons, $r$ is the distance between the ionizing source and
the illuminated edge of the cloud, $n_{H}$ is the volume density of
neutral hydrogen to ionize, and $c$ is the speed of light.  We found
that, in order to produce a \sivi/\brg~flux ratio of $\sim6$, an
ionization paramater $U$ of at least $10^{-2.7}$ is required.  When
varying the metallicity log($Z/Z_{\sun})$ from -0.5 to 1.0, we found
that the required ionization parameter varies by less than 0.3 dex.
We used $n_{H} \sim 10^{3} cm^{-3}$ as a conservative estimate of the
density in the central regions of ULIRGs~\cite[but see][whose models
suggest the densities can be as high as $10^{4}-10^{5}
cm^{-3}$]{Davies03}.  We found that the required $Q(H^{0}) \sim
4\times10^{54} (\frac{r}{750 pc})^{2} s^{-1}$.  A typical $Q(H^{0})$
for a quasar can reach nearly $10^{57} s^{-1}$~\cite[][]{Tadhunter96};
lower levels of AGN activity would naturally produce fewer ionizing
photons.  These numbers suggest that photoionization from an AGN in
the N disk could plausibly produce our measured \sivi/\brg~flux
ratios, even at a distance of 750 pc.  We note that both the N and the
SW components, at distances of 640 and 750 pc respectively, fall
within this range. However, the spatial distribution of the
\sivi/\brg~flux ratio (see \fig \ref{fig:sivibrgmap}, where the
gradient seems to be higher toward the N nucleus) is more consistent
with photoionization from a N AGN than from the SW unless there is
variable dust extinction surrounding the SE component that alters the
angle of the gradient.

We must determine if our measured \sivi/\brg~flux ratios could
instead be produced by nearby O and B stars due to intense star formation.  
To test this scenario, we used an incident O star spectrum from the
Tlusty models described in \cite{Lanz03}.  As an extreme case, we
chose two of the hottest O star spectra available, with effective
temperatures of 55,000K and  
52,500K, at solar metallicity and with log(g) of 4.5 and 4.25, respectively.  
Again we varied the ionization parameter $U$ and the metallicity of the gas to 
found the required $Q(H^{0})$ to produce our measured \sivi/\brg~flux ratios.  
In this scenario, we find that an ionization parameter $U$ of at least 
$\sim10^{12}$ is required for our measured \sivi/\brg~flux ratio.  This $U$
value is much higher than is required for an AGN because the AGN
spectrum is much harder than that of an O star.  To produce this high
of an ionization parameter from stars present within 5 pc requires 
$Q(H^{0}) \sim8\times10^{64} (\frac{r}{5 pc})^{2} s^{-1}$.  A single O star's 
$Q(H^{0}) \sim 10^{50} s^{-1}$ \cite[see models by][]{Sternberg03}; for this 
line ratio to be produced by photoionization from stars, we would need 
$10^{14}$ O stars within 5 pc -- a completely unphysical scenario.  
We conclude that if our measured \sivi/\brg~flux ratios are caused by 
photoionization, they must be from an AGN.

\subsubsection{Shock-Heated \sivi}
\label{sec:shock}

\sivi~may be ionized in gas heated by shocks and is sometimes observed in 
novae and supernova remnants \cite[][]{Gerardy01}.  Several pieces of evidence 
in our observations suggest the possibility of outflows that could be
causing shocks 
near the SE component.  The \molhy~and \feii~emission bridge, the
increase in \molhy~velocity dispersion along the minor axis, 
as well as the noted \sivi/\brg~gradient decreasing along the bridge
from the N to the SE component, would be explained by
outflows associated with the central powering source within the N nucleus 
in this scenario -- AGN jets or stellar winds.  Although the
\sivi~emission appears diffuse,  the \molhy~velocity map in
this region shows a clear shift from the systemic velocity of the
system with $\Delta v \sim -150 $km s$^{-1}$ along the bridge and 
$\Delta v \sim -75 $km s$^{-1}$ at the SE component.  The spectrum of the H
\textsc{i} 21 cm absorption shows a double-peaked profile suggesting
infall at 200 km s$^{-1}$ on scales $\leq$ 40 pc~\cite[][]{Carilli00}.
With this evidence of gas motion, it is likely that shocks may exist;
these shocks could be responsible for some or all of the measured
\sivi/\brg~flux ratio ($\leq 7.9 \pm 3.4$). 

\begin{figure*}
    \centering
    \includegraphics[width=0.8\textwidth]{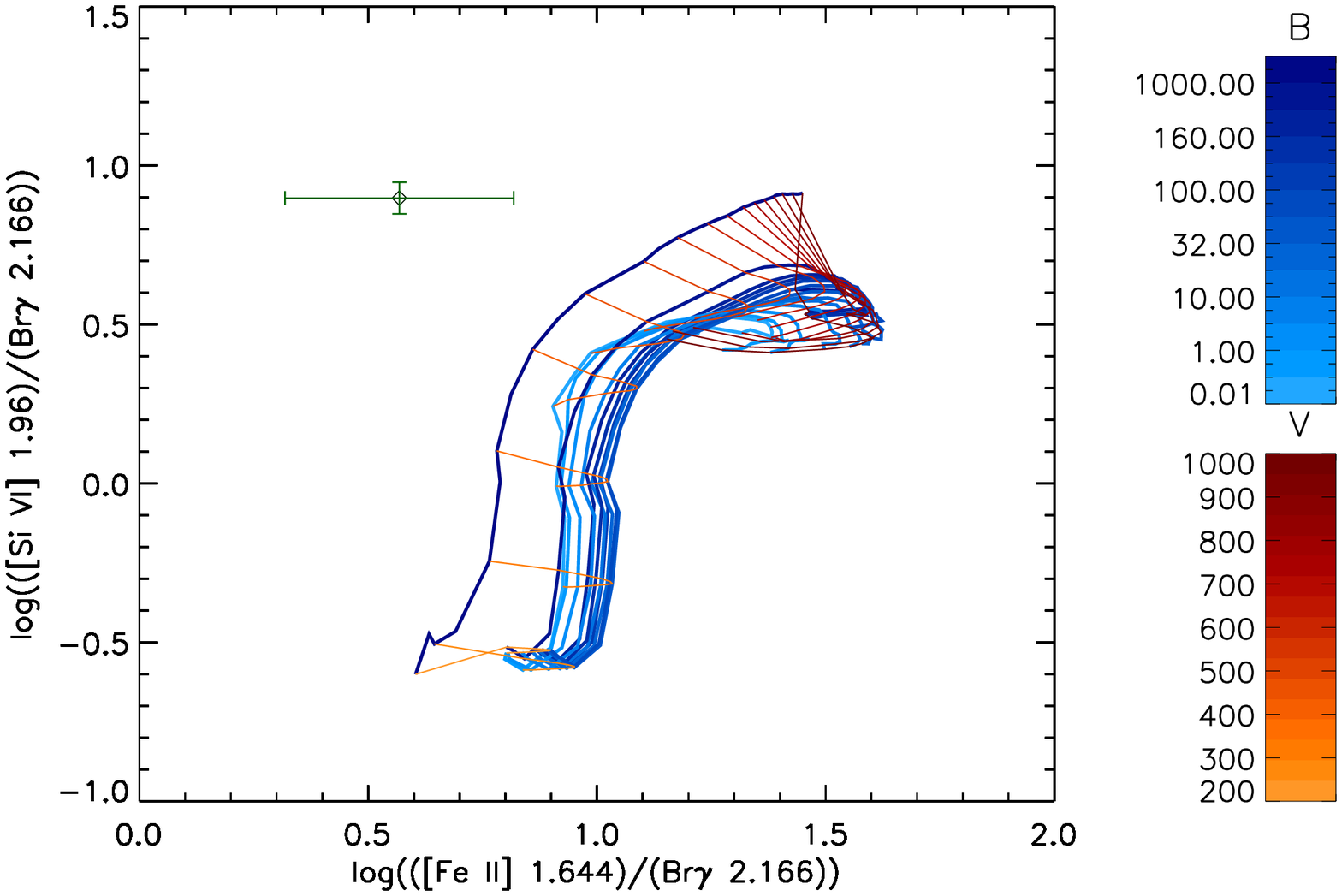}
    \caption{Diagnostic plot of \feii/\brg~vs. \sivi/\brg~showing a
      grid of shock models with solar abundance and density $n = 10^3$
      cm$^{-3}$~\cite[][]{Allen08}. The parameter space spans magnetic
      parameters $B/n^{1/2} = 10^{-4} - 10 \mu$G cm$^{3/2}$ and shock
      velocities $v_s = 200 - 1000$ km s$^{-1}$. The green symbol 
      marks the value of the line ratios associated with the SE
      component; it does not appear to converge on any shock model
      generated within our parameter space. The more robust
      \sivi/\brg~ratio (which is less sensitive to dust or cross-filter
      photometric calibrations) is only produced by models with a high
      $B/n^{1/2}$ value and high shock velocity, suggesting that only
      fast shocks in the densest material, if at all, could be
      responsible for its excitation.}   
    \label{fig:shocks}
  \end{figure*}

To determine if our observed \sivi/\brg~flux ratio may be plausibly
caused by shocks, we compared it to the shock models of~\cite{Allen08}
using the IDL SHOCKPLOT widget. In particular, we adopted the
radiative shock plus photoionized precursor model with solar abundance
and density $n = 10^3$ cm$^{-3}$ as a conservative estimate for a
ULIRG, and calculated line ratios produced for models within a grid of
parameter space.  The parameter space in magnetic parameters
$B/n^{1/2}$ and shock velocities $v_s$ of the model grid is $B/n^{1/2}
= 10^{-4} - 10 \mu$G cm$^{3/2}$ and $v_s = 200 - 1000$ km s$^{-1}$.
The model grid is shown in \fig \ref{fig:shocks}, with a 
green symbol marking the value of the line ratios observed
in the SE component; it does not appear to converge on any shock model
generated within our parameter space. The more robust \sivi/\brg~ratio
(which is less sensitive to dust or cross-filter photometric
calibrations) is only produced by models with a high
$B/n^{1/2}$ value and a shock velocity to $v_s \geq 550$ km s$^{-1}$,
indicating that only fast shocks in the densest material, if at all, could
produce the line ratios that are observed in the SE component.

\subsubsection{Energy Flow from N1}

The \sivi/\brg~ratio, seen as a gradient from the N and decreasing
toward the SE, in combination with the bridge between SE and
N seen in the other flux maps, strongly suggest an external
origin for the source of the observed energy flow, possibly from the
center of the N disk.  We find that the SE 
component is likely to be gas from a tidal feature or stellar cluster
that is being affected in one of two ways.  The easiest way to ionize
the \sivi~is with a buried AGN in the N disk or the SW AGN with variable
extinction.  The AGN would not be isotropically obscured, and thus
could photoionize gas in certain directions.  However, we find that the
observed \sivi/\brg~ratios may also be produced in extreme shock
situations.  Here we investigate if such shocks are consistent with
our kinematic data, and whether star formation is able to provide
sufficient energy to power them.  

The spectral maps and kinematics in the region of the bridge suggest
its origin as an outflow originating at the center of the N disk.  It
extends along the minor axis of the disk, following the region of
increased velocity dispersion in molecular hydrogen, suggesting that a
biconical turbulent outflow is present.  The increased \molhy~emission
toward the SE has a peculiar velocity of $-$200 km s$^{-1}$ relative
to the systemic velocity of the system; this gas appears to be outflowing.
We used simple energy arguments to estimate the amount of energy input
required, to determine whether it is likely driven by stellar winds or
from AGN feedback.  The energy contained in the outflow is $1/2~m v^2$,
where $m$ is the mass contained in the outflow and $v$ is the peculiar
velocity, $-$200 km s$^{-1}$.  To estimate the mass contained in the
outflow, we adopted solar metallicity and a density of 1000
cm$^{-3}$~\cite[a conservative estimate for (U)LIRGs,
see][]{Davies03}, and used the size of the extended emission (see
\tbl \ref{tbl:param}) assuming an equal depth and width, resulting in
a mass of 7700 $M_\odot$.  To determine an energy input rate, we
estimated the amount of time the outflow has been traveling from the
distance between the SE and the center of the N disk.  To travel this
distance of 640 pc at 200 km $s^{-1}$, the outflow requires $\sim$ 3.3
million years. Assuming the energy contained in the outflow was spread
out over this timescale, the energy input rate required to drive this
outflow is $1.3 \times 10^{43}$ erg s$^{-1}$. 

Is there enough star formation to provide this energy input in stellar
winds?  We estimate the available energy input from stellar winds as
follows: the luminosity of \brg~in the N disk is $2.1 \times 10^{40}$
erg s$^{-1}$, implying a star formation rate of 17 $M_\odot$
yr$^{-1}$~\cite[][]{Kennicutt98} and therefore a bolometric luminosity
of $1.7 \times 10^{11}~L_\odot$~\cite[][]{Krumholz07}.  Approximately
1\% of the bolometric luminosity of a starburst is available to drive
stellar winds~\cite[][]{Heckman94}, divided by 2 to account for the
collimated biconical outflow,
yielding an energy input rate available from the N disk's star
formation of approximately $3 \times 10^{42}$ erg s$^{-1}$.  
However, we note that the computed star formation rate may be
underestimated since \brg~in the N disk may be subject to dust
extinction. There may also exist an uncertainty of a factor of a few
in the luminosity of starburst-driven winds.  Thus, although the
energy from star formation seems insufficient by a factor of 4
compared to the required energy input rate, starbursts cannot be
definitively ruled out as a source of driving the winds.  

It is also important to consider whether the energy available to drive
outflows from star formation is sufficient to shock-heat the
\sivi~seen in the SE component.  The incoming kinetic energy rate is 
$e_k = 1/2~\rho v_s^3A$, where $\rho$ is the density (as above, 1000
cm$^{-3}$), $v_s$ is the shock velocity ($\geq$ 550 km s$^{-1}$,
calculated from the SHOCKPLOT models in \S \ref{sec:shock}), and $A$
is the surface area of the gas being shocked. A lower limit on the
shocked area is a hemisphere on the surface of the SE component: 2$\pi$
(70 pc)$^2$.  (In reality, the shock front would likely have
structure and therefore larger surface area; this provides a
conservative lower limit on the energy input required to shock this
\sivi.)  These estimates require an energy input of $1.7 \times
10^{44}$ ergs$^{-1}$, two and a half orders of magnitude larger
than $5 \times 10^{41}$ erg s$^{-1}$, the available energy for
outflows from star formation in the N disk. 
This suggests that the observed \sivi~emission cannot be caused solely
by shock-heating due to stellar winds.  Stronger shocks may be
plausible with AGN-driven outflows, but because shock models cannot
produce the observed \sivi/\brg~ratio, at least some of the
\sivi~emission must be caused by AGN photoionization 
(as discussed in \S \ref{sec:photo}). 

Though photoionization from an AGN is the simplest way to produce the
\sivi~emission in the SE, it is likely produced by a combination of
the two processes: e.g., a 
weaker AGN to photoionize some gas, and less extreme shocks to heat
the rest.  Our observed double-peaked \sivi~velocity profile indicates
that there are likely two kinematic components in this region --- with
broad and narrow velocity dispersions, respectively.  The 
broad component contributes roughly half the \sivi~flux, resulting
in a \sivi/\brg~ratio of 3.9$\pm$1.7, or 0.59$^{+0.16}_{-0.25}$ dex in
logarithmic units. This constrains the shock models to velocities $v_s
\sim 350-450$ km s$^{-1}$, within the range of the velocity dispersion
of the broad component.  The two narrow and broad components may have
originated from two different heating mechanisms, 
lending support to the combined AGN and shocks scenario. 
A simple explanation might involve a combination of shocks from the
outflow and photoionization by a buried AGN within the N disk.

The SW AGN is also a potential origin for the photoionizing flux, even
though it is further in projection from the SE component.  In this
alternative scenario, the observed spatial gradient in the
\sivi~excitation toward the N nucleus could be caused by differential
extinction in the dust between the SW and the SE components,
by differential gas densities within the SE component, or by geometric
projection effect.  Current observations are unable to distinguish between
these proposed scenarios, and thus further investigations will be
needed to confirm the nature of the ionizing source within Mrk 273's nucleus.

  \section{Summary}
  \label{Summary}

  High spatial resolution imaging and integral field spectroscopy with 
  Keck laser guide star adaptive optics has allowed us to study the
  properties of the atomic and molecular emission lines of Mrk 273 in
  the $H$- and $K'$-bands.  The nuclear region is resolved into three
  primary components (N, SW, and SE) with a gas bridge between the N
  and SE components.  
  These near-infrared high-resolution maps are well suited for
  investigating the gas dynamics and the location of the hardest
  photons from AGN and fast shocks; they complement other
  multiwavelength data for determining the ultimate power associated
  with each component.  

  We have shown that the N component is a rotating gaseous disk with
  PA $\sim 330^\circ$. The characteristic size of the disk varies
  depending on the emission line studied, but the effective radius is
  roughly $0\farcs30$ (240 pc). Using the observed strong rotation, we
  measure from the kinematics of \feii~a central mass of $M = 1.04 \pm
  0.1 \times 10^9 M_\odot$, consistent with the mass measurement from
  OH maser kinematics~\cite[][]{Klockner04}.  
 This enclosed mass likely indicates the presence of a supermassive 
 black hole at the center of the N disk, suggesting that the N
 component is the remnant nucleus of a progenitor galaxy.  The gaseous
 disk may be a remnant from the  
  progenitor galaxy or may be recently formed with new gas from the
  merger.  The black hole is heavily obscured and may be an AGN and/or
  surrounded by additional star formation, consistent with conclusions from 
  X-ray~\cite[][]{Iwasawa11_mrk273} and radio data~\cite[][]{Carilli00,Bondi05}. 
  An increase in the velocity dispersion of \molhy~along the minor
  axis of the N disk and toward the SE suggests the presence of
  bipolar molecular outflow originating from the N source. Extended
  emission of shocked \molhy~toward the SE also reveals the presence
  of a collimated outflow.  Energy arguments associated with the size
  and kinematics of this outflow suggest that it is too massive to be
  produced by star formation alone, and also point toward the
  scenario in which the N black hole is a buried AGN.  
  \molhy~line ratios offer
  insight into the rotational and temperature conditions as well as
  their excitation mechanisms.  We derived $T_{\rm vib} \simeq
  2591\pm79$K and $T_{\rm rot} \simeq 1604\pm113$K, which fall within
  the range of temperatures for Seyfert
  galaxies~\cite[][]{Reunanen02}.  

  The SW component is bright and compact in the $K'$- and $H$-band
  continua, and shows emission of the coronal line \sivi.
  We suggest that a non-thermal photon source is ionizing the
  silicon and producing the \sivi~emission that we
  observe; it would also heat the hot dust that we observe in the continuum.
  Our near-infrared data support the notion that the SW nucleus hosts the
  known hard X-ray AGN in this system~\cite[][]{Iwasawa11_mrk273}.  

  The SE component presents a new piece in the puzzle of the merger's
  nuclear region.  While it has previously been detected in the radio
  regime~\cite[][]{Smith98}, it has not been detected in optical and
  near-infrared images until now due to the dominance of the broadband continuum.
  With our new AO-assisted NIRC2 images and OSIRIS integral field
  spectroscopy, the SE component is detected in several emission
  lines, and is prominent in both \feii~and \sivi. The double-Gaussian
  fit to this \sivi~line indicates that there are two velocity
  components involved, 
  perhaps due to outflows.  We discussed several potential scenarios
  for the origin of the SE component: photoionization by AGN or O and
  B stars, shock-heating due to stellar winds or AGN-driven outflows,
  or a combination of both. We favor the case 
  where the SE component is a clump of gas or tidal feature
  photoionized by an obscured AGN in the N disk and/or shock-heated by
  outflows from the same AGN and its surrounding star formation. 
  Alternatively, the photoionizing flux affecting the SE component may
  originate from the SW AGN.  Confirming the N black hole as a buried
  AGN and the nature of the source for the SE component's ionizing
  radiation will require further investigations. 
  While AGN photoionization alone could produce the observed
  \sivi/\brg~ratio, only the most extreme shocks in the shock-heated
  scenario could account for these \sivi/\brg~values. We suggest that
  a combination of photoionization and shocks could best produce both
  the observed line ratios and the two kinematic components of \sivi. 
 
  As a late-stage gas-rich merger, Mrk 273  offers a unique time
  in the merger sequence where the nuclei have not completely merged
  and yet one AGN has already been triggered.  We have identified a second 
  black hole, located in the center of the N disk, which may be an 
  obscured AGN with supporting evidence of outflows.  The gas disk is
  likely the remnant nuclear disk from one of the progenitor galaxies,
  or perhaps a young disk formed during the merger. 
  An  understanding of the unveiling of this second black hole as an
  unburied AGN may lead to a deeper understanding of the frequency of
  dual AGNs in merging gas-rich galaxies.

\acknowledgements

We thank the anonymous referee for a helpful discussion on
alternative cases for photoionizing the SE component, and for various
suggestions on strengthening the scenarios put forth in the paper.  
We acknowledge H. Inami and S. Stierwalt for providing their
\emph{Spitzer}-IRS spectra toward ease of target selection for our
observing program; M. Ammons for help with data reduction and PSF
estimation; R. da Silva for help with CLOUDY and shock-plot modeling;
and D. Rupke for helpful conversations toward the results and their
interpretations. We also thank the Keck staff for help
with carrying out the observations and the UH/UC TACs for granting
this observing time.   
The data presented herein were obtained
at the W.M. Keck Observatory, which is operated as a scientific
partnership among the California Institute of Technology, the
University of California and the National Aeronautics and Space
Administration. The Observatory was made possible by the generous
financial support of the W.M. Keck Foundation.  
The authors wish to recognize and acknowledge the very significant
cultural role and reverence that the summit of Mauna Kea has always
had within the indigenous Hawaiian community.  We are most fortunate
to have the opportunity to conduct observations from this mountain. 
VU acknowledges funding support from the NASA Harriet G. Jenkins
Predoctoral Fellowship Project and the Smithsonian Astrophysical
Observatory Predoctoral Fellowship. AM acknowledges funding support
from the NSF Graduate Fellowship. KI thanks support from Spanish
Ministerio de Ciencia e Innovaci\'on (MICINN) through grant
(AYA2010-21782-C03-01). This work is partially supported 
by the JPL Contract/IRAC GTO Grant \# 1256790.

\bibliography{thesis}
\bibliographystyle{apj}

\end{document}